\documentclass[]{spie}  
 
\usepackage{amsmath,amsfonts,amssymb}
\usepackage{graphicx}
\usepackage[colorlinks=true, allcolors=blue]{hyperref}
\usepackage{array}
\newcolumntype{P}[1]{>{\centering\arraybackslash}p{#1}}
\def\cryoac{cryo-AC}
\def\xisc{X-ISC}

\title{The Athena X-ray Integral Field Unit: a consolidated design for the system requirement review of the preliminary definition phase}
\begin{document} 
\authorinfo{Further author information: (Send correspondence to Didier Barret \& Philippe Peille)\\Didier Barret: E-mail: dbarret@irap.omp.eu, Philippe Peille: E-mail Philippe.Peille@cnes.fr}
\author[1]{Didier Barret}
\author[2]{Vincent  Albouys}
\author[3]{Jan-Willem den Herder}
\author[4]{Luigi Piro}
\author[5]{Massimo Cappi}
\author[6]{Juhani Huovelin}
\author[7]{Richard Kelley}
\author[8]{J. Miguel Mas-Hesse}
\author[9]{Stéphane Paltani}
\author[10]{Gregor Rauw}
% ...
\affil[1]{Institut de Recherche en Astrophysique et Planétologie, Université de Toulouse, CNRS, UPS, CNES 9, Avenue du Colonel Roche, BP 44346, F-31028, Toulouse Cedex 4, France}
\affil[2]{Centre National d'Etudes Spatiales, Centre spatial de Toulouse, 18 avenue Edouard Belin, 31401 Toulouse Cedex 9, France}
\affil[3]{SRON, Netherlands Institute for Space Research, Niels Bohrweg 4, 2333 CA Leiden, The Netherlands}
\affil[4]{Istituto di Astrofisica e Planetologia Spaziali, Via Fosso del Cavaliere 100, 00133, Roma, Italy}
\affil[5]{INAF, Osservatorio di Astrofisica e Scienza dello Spazio, via Gobetti 93/3, 40129, Bologna, Italy }
\affil[6]{Department of Physics, Faculty of Science, P.O. Box 64 (Gustaf Hällströmin katu 2), FI-00014, University of Helsinki, Finland}
\affil[7]{NASA Goddard Space Flight Center, 8800 Greenbelt Rd, Greenbelt, MD 20771, United States}
\affil[8]{Centro de Astrobiología (CSIC-INTA), Dep. de Astrofísica, Unidad María de Maeztu, European Space Astronomy Centre (ESA-ESAC), Camino Bajo del Castillo s/n - Villafranca del Castillo, 28692 Villanueva de la Cañada, Madrid, Spain}
\affil[9]{Département d'Astronomie, Université de Genève, Chemin d’Ecogia 16, CH-1290 Versoix, Switzerland}
\affil[10]{Université de Liège, Institut d'Astrophysique et de Géophysique, Quartier Agora, Allée du 6 Août 19c, B-4000 Liège 1 (Sart-Tilman), Belgium}
% ...

\maketitle

\newpage
\begin{abstract}
The Athena X-ray Integral Unit (X-IFU) is the high resolution X-ray spectrometer studied since 2015 for flying in the mid-30s on the Athena space X-ray Observatory. Athena is a versatile observatory designed to address the Hot and Energetic Universe science theme, as selected in November 2013 by the Survey Science Committee. Based on a large format array of Transition Edge Sensors (TES), X-IFU aims to provide spatially resolved X-ray spectroscopy, with a spectral resolution of 2.5 eV (up to 7 keV) over an hexagonal field of view of 5 arc minutes (equivalent diameter). The X-IFU entered its System Requirement Review (SRR) in June 2022, at about the same time when ESA called for an overall X-IFU redesign (including the X-IFU cryostat and the cooling chain), due to an unanticipated cost overrun of Athena. In this paper, after illustrating the breakthrough capabilities of the X-IFU, we describe the instrument as presented at its SRR (i.e. in the course of its preliminary definition phase, so-called B1), browsing through all the subsystems and associated requirements. We then show the instrument budgets, with a particular emphasis on the anticipated budgets of some of its key performance parameters, such as the instrument efficiency, spectral resolution, energy scale knowledge, count rate capability, non X-ray background and target of opportunity efficiency. Finally, we briefly discuss the ongoing key technology demonstration activities, the calibration and the activities foreseen in the X-IFU Instrument Science Center, touch on communication and outreach activities, the consortium organisation and the life cycle assessment of X-IFU aiming at minimising the environmental footprint, associated with the development of the instrument. Thanks to the studies conducted so far on X-IFU, it is expected that along the design-to-cost exercise requested by ESA, the X-IFU will maintain flagship capabilities in spatially resolved high resolution X-ray spectroscopy, enabling most of the original X-IFU related scientific objectives of the Athena mission to be retained. \\
{\it The X-IFU will be provided by an international consortium led by France, The Netherlands and Italy, with ESA member state contributions from Belgium, Czech Republic, Finland, Germany, Poland, Spain, Switzerland, with additional contributions from the United States and Japan.}
\end{abstract}


\newpage
\section{X-IFU: a breakthrough capability for Athena}
We first recap the way X-IFU will address its core scientific objectives, emphasizing on the drivers of its performance requirements, which are met in its current design.
\subsection{To address its core scientific objectives}

ESA’s Athena X-ray observatory mission was selected in 2014 to address the Hot and Energetic Universe science theme \cite{2013arXiv1306.2307N,Barret_2013sf2a.conf..447B,barcons2015JPhCS.610a2008B,Barcons2017AN....338..153B,Barret2020AN....341..224B,Pointecouteau2013arXiv1306.2319P,Ettori2013arXiv1306.2322E,Croston2013arXiv1306.2323C,Kaastra2013arXiv1306.2324K,Aird2013arXiv1306.2325A,Georgakakis2013arXiv1306.2328G,Cappi2013arXiv1306.2330C,Dovciak2013arXiv1306.2331D,Branduardi2013arXiv1306.2332B,Sciortino2013arXiv1306.2333S,Motch2013arXiv1306.2334M,Decourchelle2013arXiv1306.2335D,Jonker2013arXiv1306.2336J}. The Hot Universe refers to the baryons in the Universe at temperatures above $10^{5-6}$ K, which accounts for most of its baryons content. The Energetic Universe refers to all phenomena occurring in the vicinity of compact objects across the mass spectrum. Supermassive black holes, for instance largely influence their surroundings out to large-scales through a poorly understood process called Cosmic Feedback. As X-rays are copiously produced by hot gas and accretion around black holes, the best observational handle on the Hot and Energetic Universe is through X-ray observations. In addition, X-rays can escape relatively unimpeded from significantly obscured environments and are sensitive, for a wide range of column densities, to all ionization states (from cold to highly ionized gas) of all most abundant elements (such as C, N, O, Si, Mg, Fe, Ni, etc.). 
\begin{figure}[!b]
    \centering
    \includegraphics[width=16.5cm]{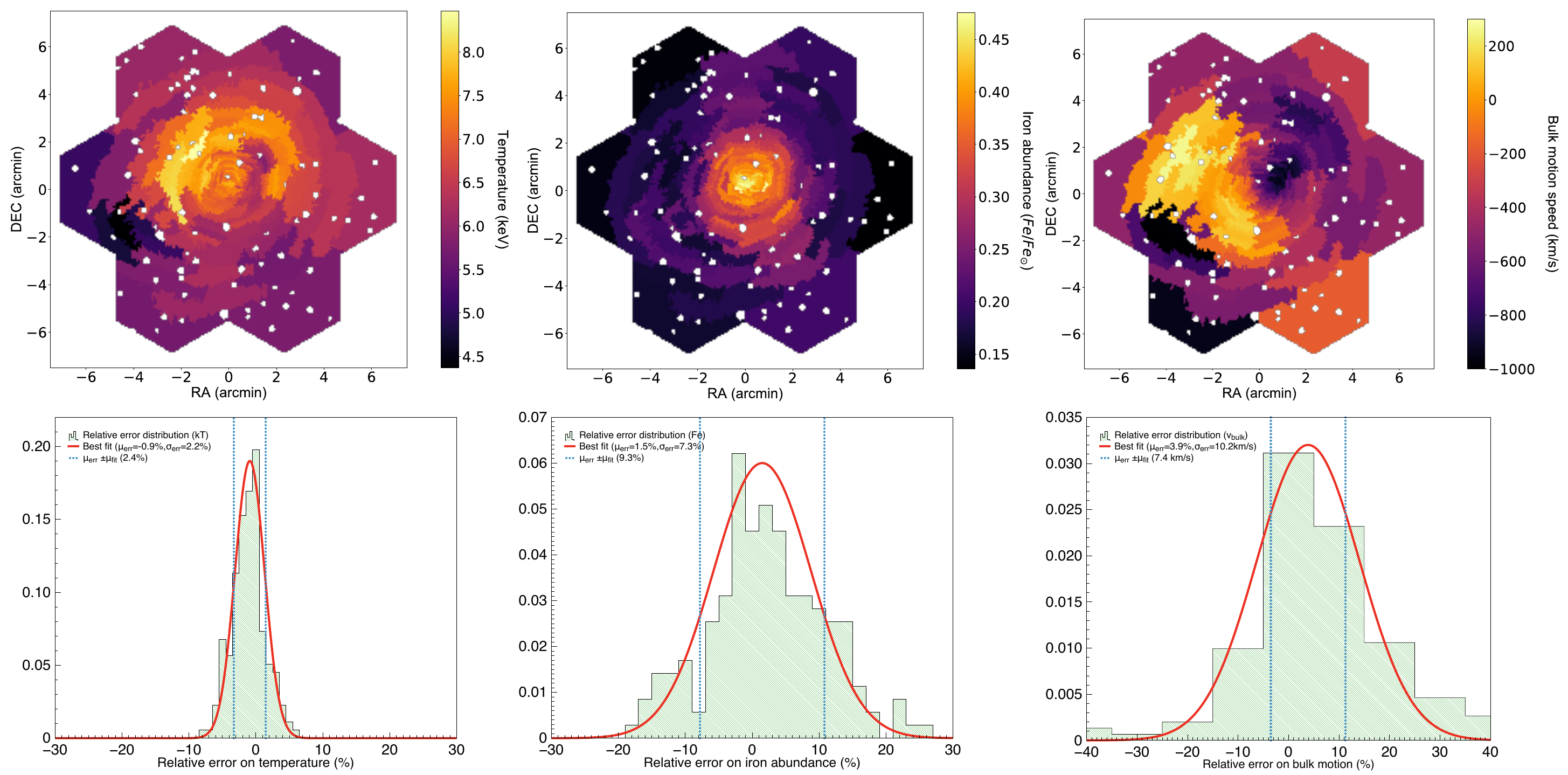}
    \caption{Examples of (Top:) Spatial distribution profiles of the ICM parameters for a nearby "typical" cluster. (From left to right): Plasma  temperature in keV, abundance of iron (with respect to solar) and bulk motion (km/s) computed from the mean redshift of the cluster. White dots represent the excised point sources. (Bottom:) Corresponding normalised relative error distribution (\%) of output with respect to input histogram (green) along with Gaussian best fit (red solid line). Blue dashed lines indicate the mean values of the statistical error related to the fit.
    (Credits: \cite{cucchetti2018A&A...620A.173C}, simulations courtesy: S. Borgani, V. Biffi, K. Dolag, L. Tornatore, E. Rasia - INAF Trieste) }
    \label{fig:cluster}
\end{figure} 
To address its science objectives, Athena has been conceived as a large observatory requiring in particular the capability to perform spatially-resolved high-resolution spectroscopy, as well as the capability to observe bright X-ray sources, including also its fast repointing capability. This defines the sizing requirements for the Athena X-ray Integral Field Unit (X-IFU) \cite{2018SPIE10699E..1GB}. 

In short, the X-IFU will enable for the first time, critical observations of the hot and energetic universe, in particular 1) the study the chemical evolution of the Universe along with the physical processes governing the assembly and the evolution of the large-scale structures, and 2) the determination of how black holes (and all types of compact objects) work and shape the Universe. The breadth of the science affordable with the X-IFU will encompass key scientific issues of the Hot and Energetic Universe science theme and beyond. 

X-IFU will provide revolutionary observations through: 

\begin{itemize}
  \item Integral field spectroscopic mapping of hot cosmic plasmas, enabling 3-D measurements of gas bulk motions and turbulence, chemical abundances and the spatial distribution of these and other physical parameters in local and distant galaxy clusters (see Figure \ref{fig:cluster}). This drives the X-IFU field of view and spatial resolution, quantum efficiency, particle background, spectral resolution and calibration accuracy. 
\end{itemize}
\begin{figure}[!h]
    \centering
    \includegraphics[width=8.5cm]{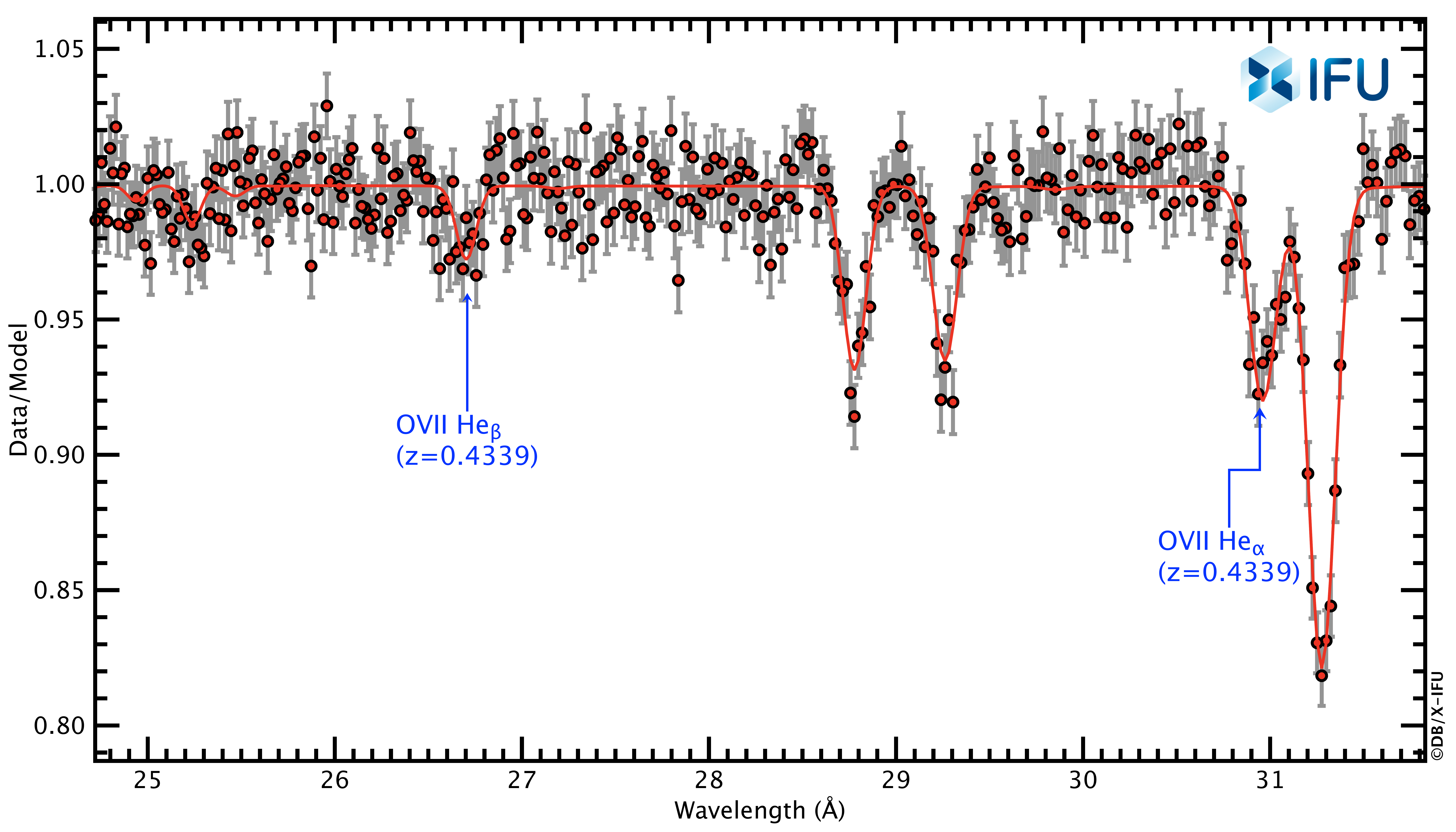}\includegraphics[width=8.5cm]{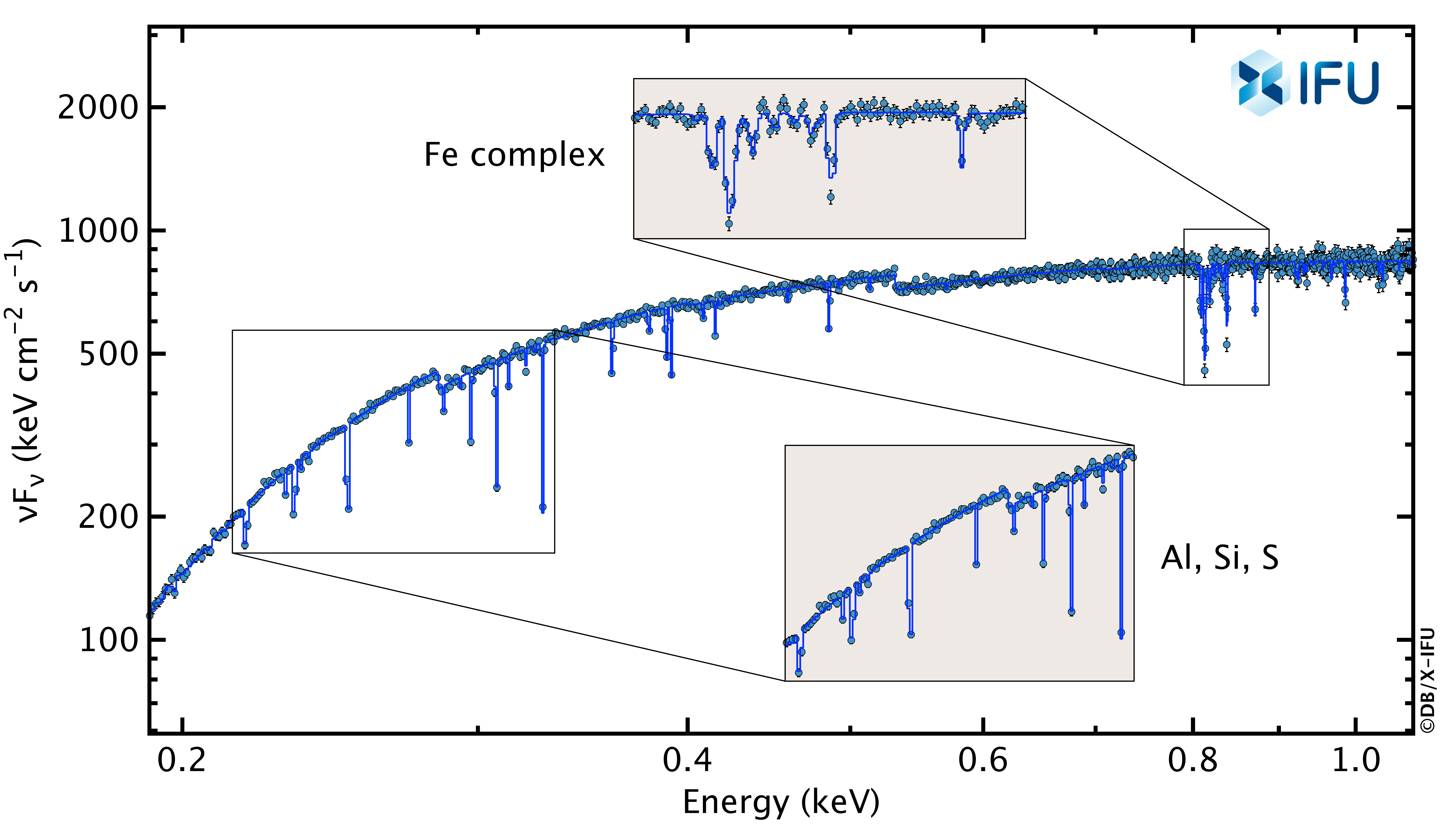}
    \caption{Left) Simulated X-IFU spectrum of an intervening absorber  in the Warm-Hot Intergalactic Medium at z=0.4339. The plot shows the ratio of the X-IFU data of the blazar 1ES 1553+113 with the local best-fitting continuum model, to highlight the two OVII He-$\alpha$  and He-$\beta$ absorption lines. The simulation is performed using the best fitting parameters derived from the long XMM-Newton RGS observations\cite{nicastro2018Natur.558..406N}. Right) A simulated X-IFU X-ray spectrum of a medium bright (fluence $=0.4 \times 10^{-6}$ erg cm$^{-2}$) Gamma-Ray Burst (GRB) afterglow at $z = 7$, characterized by deep narrow resonant lines of Fe, Si, S, Ar, Mg, from the gas in the environment of the GRB. An effective intrinsic column density of $2 \times 10^{22}$ cm$^{-2}$ has been adopted. }
    \label{fig:whim}
\end{figure} 
\begin{itemize}
\item High sensitivity to line detection, enabling the detection of absorption and emission lines from Warm and Hot Intergalactic Medium (WHIM) filaments, and weak spectral features produced by unusual ion species or states (see Figure \ref{fig:whim}). This drives the X-IFU spectral resolution, quantum efficiency, calibration, throughput, and fast reaction time. 

\begin{figure}[!h]
    \centering
    \includegraphics[width=8.5cm]{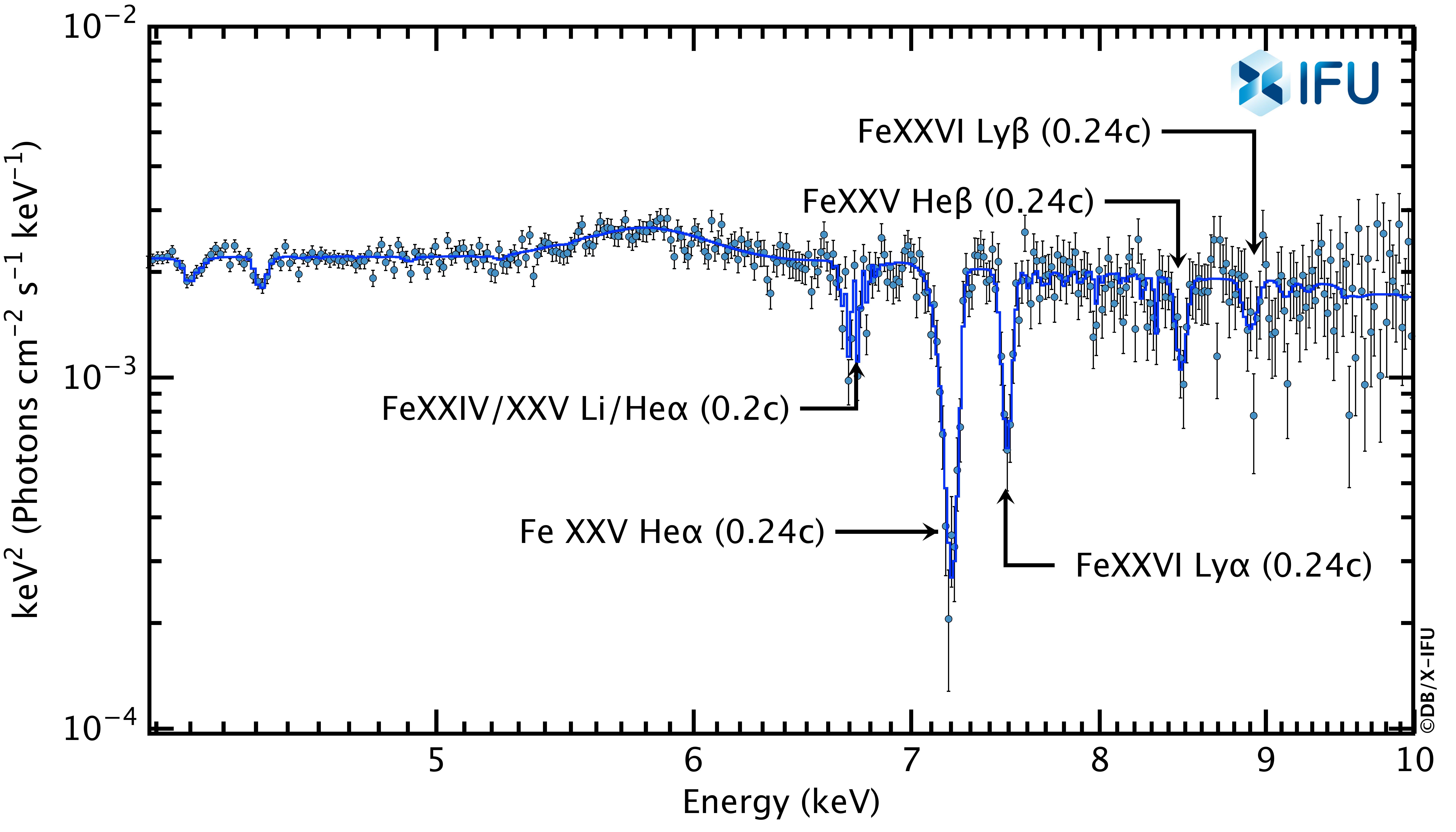}\includegraphics[width=8.5cm]{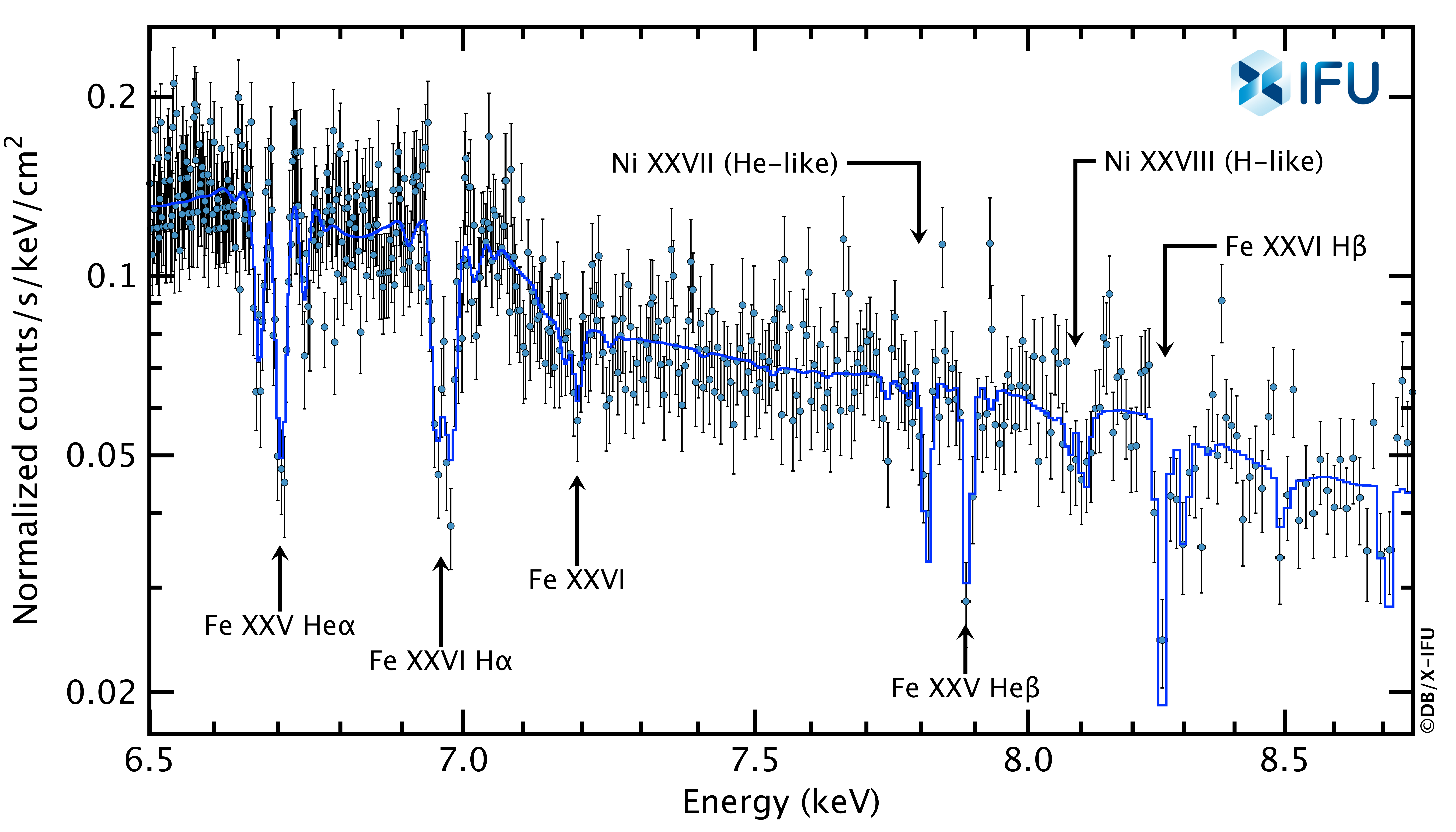}
    \caption{Left) Probing (slow) warm absorbers and (fast) massive outflows in AGN. Simulated 100 ks X-IFU spectrum of PDS456 (z=0.184) above 4 keV obtained assuming a source state as in the XMM+NuSTAR observations of PDS456 as shown in \cite{reeves2018ApJ...867...38R}. Right) X-IFU simulated observation lasting only ~120 seconds of the Black Hole binary GRS1915+105. A disk wind, as reported in \cite{miller2016ApJ...821L...9M} has been simulated. Strong spectral features can be clearly seen in the spectrum.This rich set of features will enable unprecedented studies of the structure of the disk winds.}
    \label{fig:pds}
\end{figure} 

\item Physical characterization of the Hot and Energetic Universe, including plasma diagnostics using emission line multiplets, absorption lines (see Figure \ref{fig:pds}), AGN reverberation and black hole spin measurements, winds in galactic sources in outburst, AGN winds, and outflows. This drives the X-IFU spectral resolution, calibration, high-count rate capability, and fast reaction time.
\end{itemize}

\subsection{To surpass all other past and future instruments/missions}
\begin{figure}[!h]
    \centering
    \includegraphics[width=16cm]{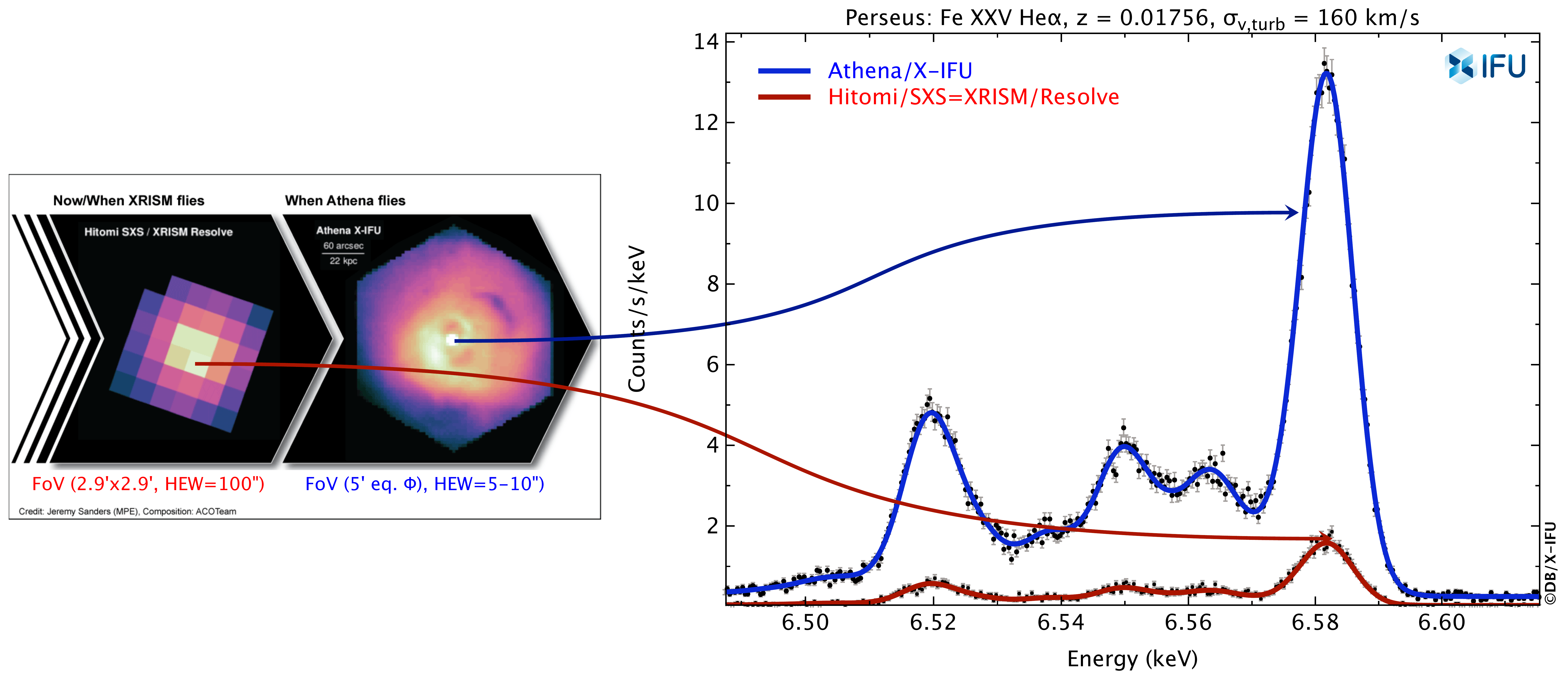}
    \includegraphics[width=8.5cm]{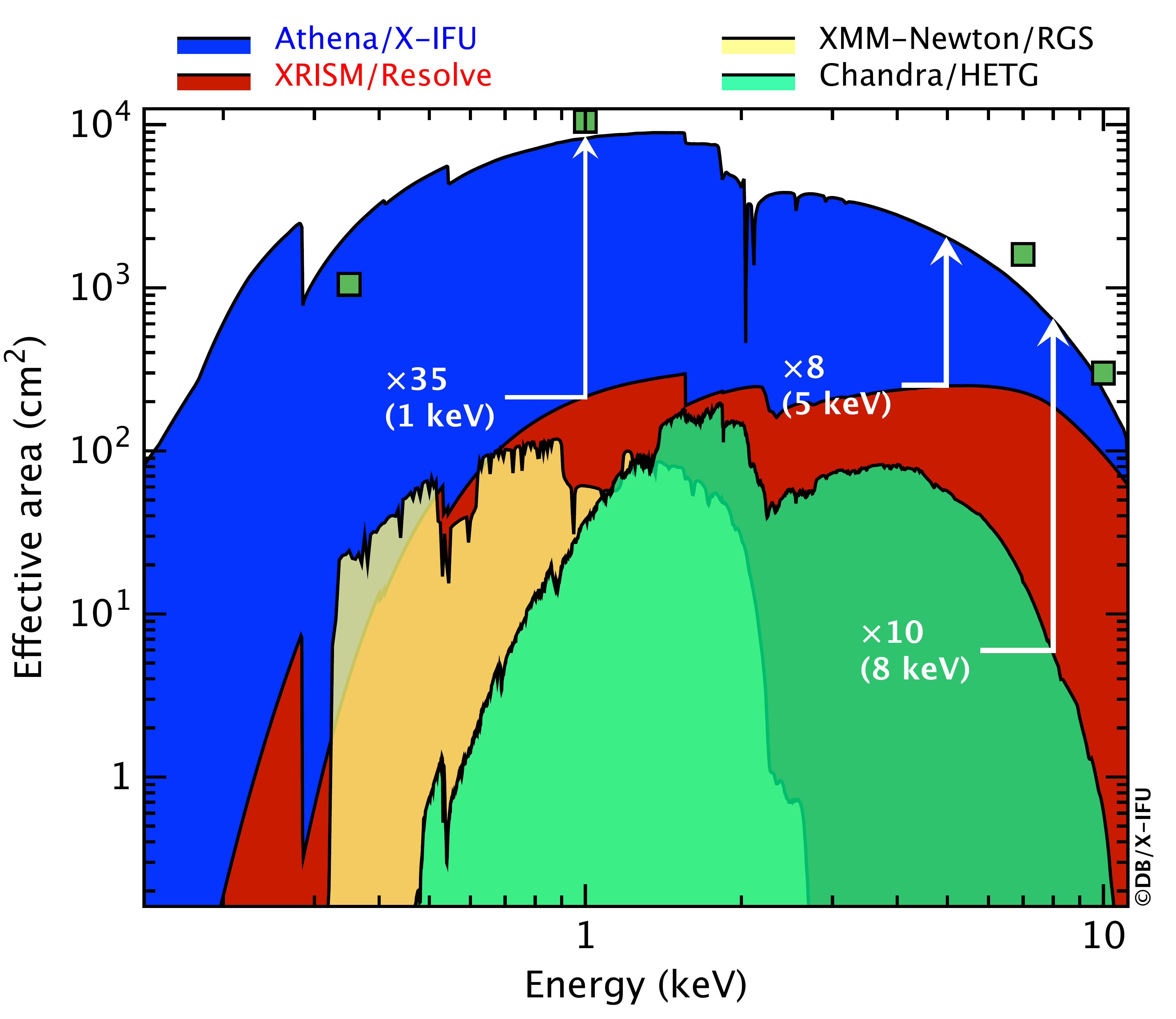}\includegraphics[width=8.5cm]{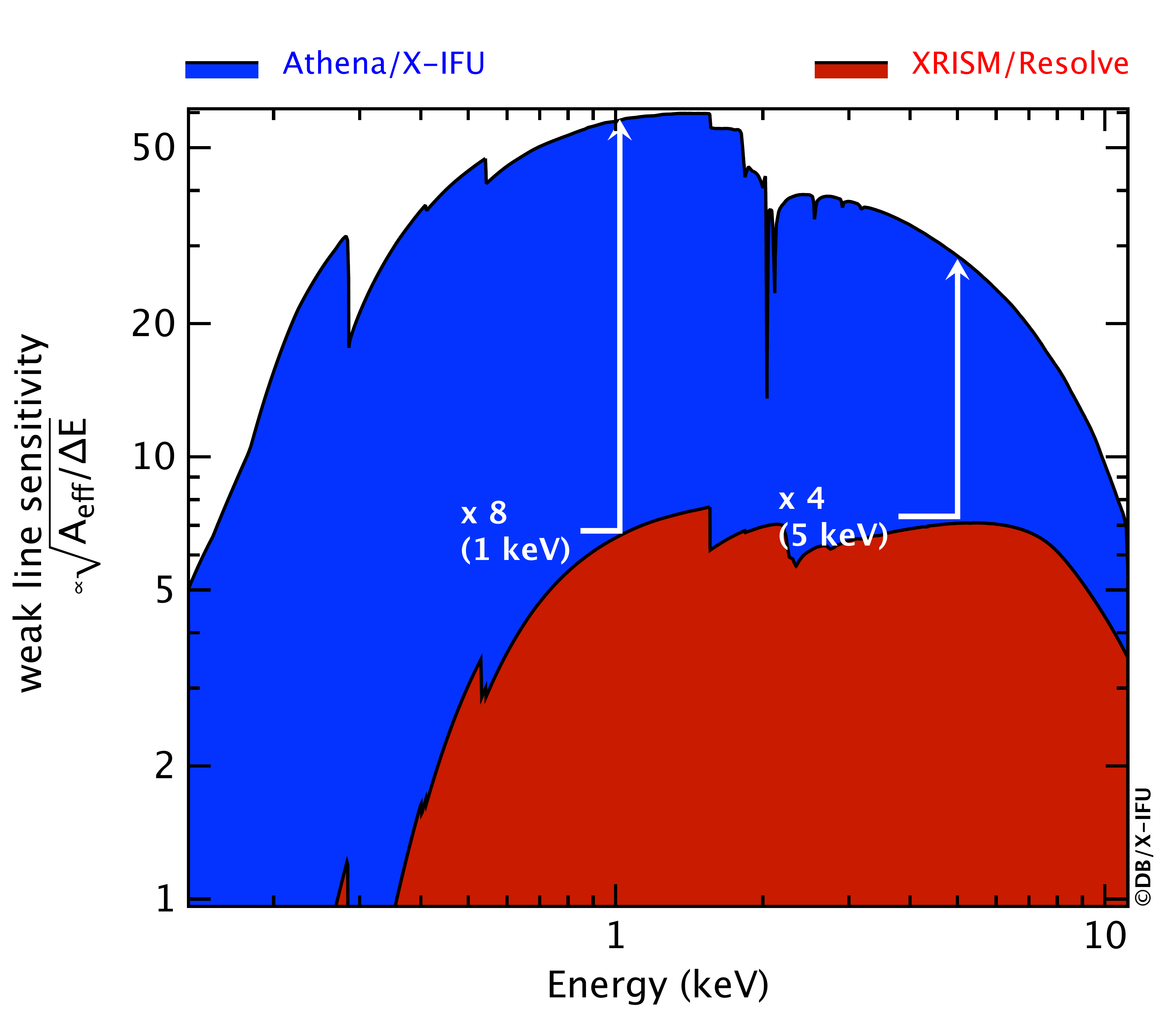}
    \caption{Top) The Hitomi/SXS=XRISM/Resolve and X-IFU images and spectra of Perseus cluster. Credit: Jeremy Sanders (MPE). This combined figure shows at the same time the significant improvement in fine imaging, over a comparable field of view, as well as the higher throughput of X-IFU compared to Resolve (the two spectra are simulated with the same exposure time). Bottom) Effective area (left) and weak narrow line sensitivity (right) of X-IFU in comparison with the one of XRISM/Resolve and Chandra/XMM-Newton gratings (not encapsulated in that figure, the fact that the angular resolution will be ten times better for Athena/X-IFU compared to XRISM/Resolve). The effective area requirement applying to X-IFU are indicated with filled green square symbols. This assumes the mirror telescope configuration as described in ATHENA - Telescope Reference Design
and Effective Area Estimates, ESA-ATHENA-ESTEC-PL-DD- 001, Issue 3.3, 21/12/2020 and the instrument efficiency presented in \S \ref{subsec_ie}.}
    \label{fig:fom_image}
\end{figure}

In the current X-ray science context, the X-IFU has been thought and designed to provide unprecedented spatially-resolved spectroscopy with performances greatly exceeding those offered by current X-ray observatories like XMM-Newton and Chandra, or XRISM with its Resolve high resolution spectrometer \cite{Guainazzi2020IAUS..342...29G}. This capability comes from an X-ray telescope combining unprecedented collecting area (1.4 m$^2$ at 1 keV), a very good angular resolution (5-10”) to provide sufficient counts on a 5' (equivalent diameter) field of view, a better than $2.5$ eV energy resolution below 7 keV, and a sensitive band pass going from 0.2 to 12 keV (see Figure \ref{fig:fom_image})

\section{The X-IFU instrument}
The X-IFU key performance requirements matching the Athena science requirements are listed in Table \ref{tab:performance}.

\begin{table}[!h]
    \centering
    \begin{tabular}{|l|l|}
\hline
Key performance parameter &	Value \\
\hline
Energy range &	0.2-12 keV \\
Spectral resolution &	$<$ 2.5 eV up to 7 keV \\
Energy scale calibration &	0.4 eV in the 0.2-7 keV range \\
Field of view&	5’ (equivalent diameter) \\
Instrument efficiency at 0.35, 1.0, 7.0, 10 keV &	$>13$\%, $>57$\%, $>63$\%, $>42$\% \\
Non X-ray background&	$< 5\times 10^{-3}$ counts/s/cm$^2$/keV (E $> 2$ keV) \\
Relative time resolution&	10 $\mu s$ \\
2.5 eV throughput (broadband, point source)	&80\% at 1 mCrab (goal of 10 mCrab) \\
10 eV throughput (5-8 keV, point source)&	50\% at 1 Crab \\
2.5 eV throughput (broadband, extended source)	&80\% at $ 2\times10^{-11}$ ergs/s/cm$^2$/arcmin$^2$ \\
Continuous cool time / regeneration time&	Up to 28.5 hours (75\% duty cycle)\\
\hline
    \end{tabular}
    \caption{X-IFU top level performance requirements.}
    \label{tab:performance}
\end{table}

\begin{figure}[!h]
    \centering
    \includegraphics[width=17cm]{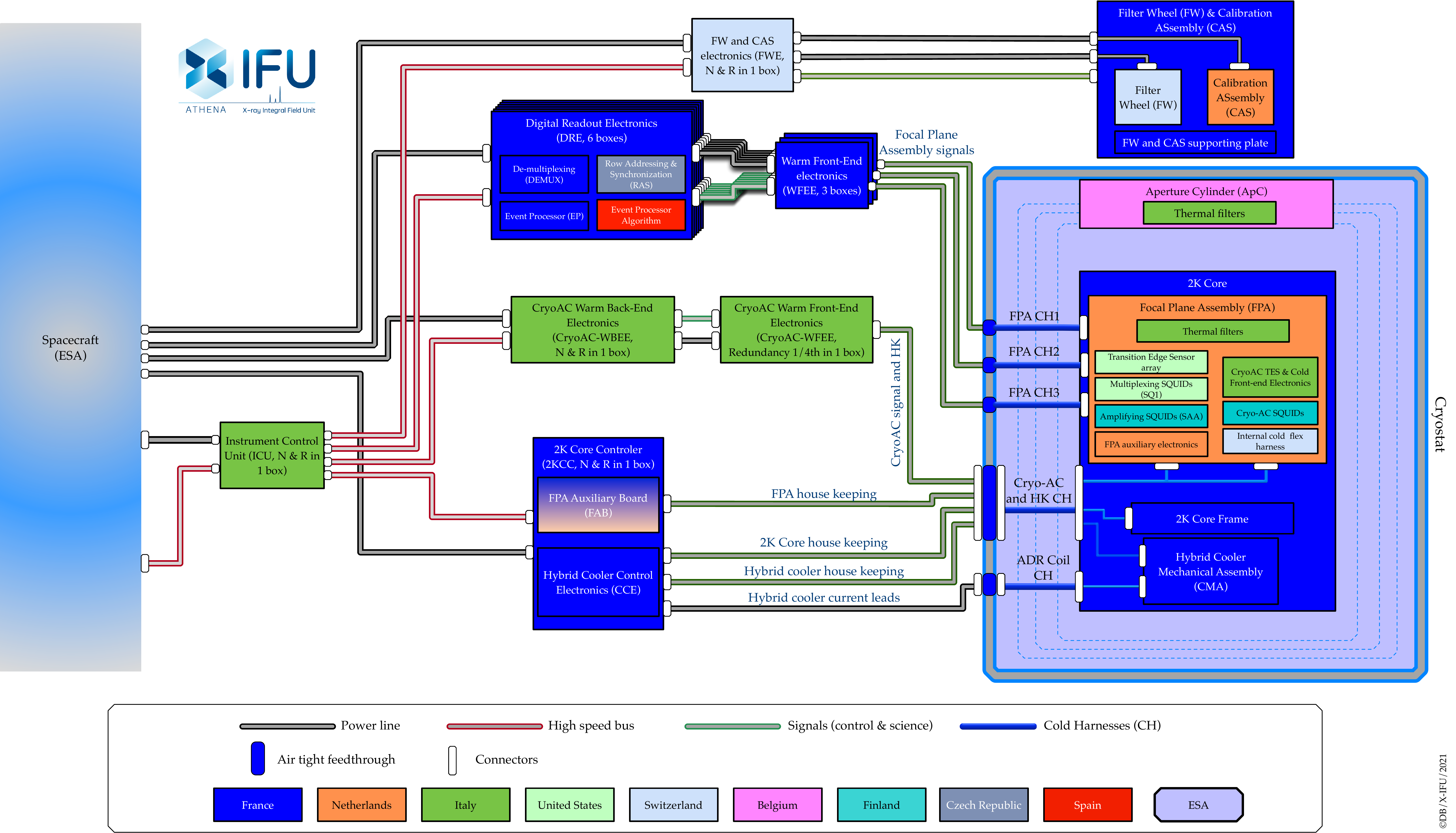}
    \caption{The Physical Breakdown of the X-IFU instrument, highlighting its main components, as well as the country responsible for the procurement.}
    \label{fig:phy_breakdown}
\end{figure}

Next, we move to the description of the instrument, whose physical and programmatic breakdown is shown in Figure \ref{fig:phy_breakdown}. In the following sections, we will go through all the main components of the instrument. It is hosted by the Athena Science Instrument Module, which provides in particular the cryostat and the associated cooling chain. These are not described in this paper (see\cite{Barret_2013arXiv1308.6784B,Ravera_2014SPIE.9144E..2LR,Barret_2016SPIE.9905E..2FB,Barret_2018SPIE10699E..1GB,Pajot_2018JLTP..193..901P} for a description of the previous incarnations of X-IFU, when the cryostat and cooling chain were in the Consortium perimeter, and still relevant for the global SIM architecture today.).

\subsection{The Transition Edge Sensor (TES) Array}
\label{sec:tes_array}
TESs are superconducting thin film thermistors operated in the sharp transition region between the superconducting and normal states. The X-IFU TESs consist of a 50 $\mu$m-square Mo/Au bilayer voltage biased at a superconducting-to-normal transition temperature of $\sim 90$ mK. The TES is deposited atop a 0.5 $\mu$m-thick silicon-nitride membrane, which forms a weak thermal link to the heat sink. TESs are coupled to an electroplated Au/Bi/Au absorber, which provides the X-ray stopping power. Its composition (Bi to Au ratio) is chosen to provide the desired pixel heat capacity (optimized for energy resolution and dynamic range), while simultaneously achieving the desired pixel quantum efficiency. This optimization has a limit, as a minimal layer of gold is required for intra-pixel heat diffusion. A small gold cap is deposited on top of the absorber to optimize infrared reflectivity and protect the porous bismuth from humidity. For the baseline 317 $\mu$m pitch pixels, a target of $\sim 91$\% intrinsic stopping power at 7 keV is achieved with a total gold thickness of 1.09 $\mu$m (of which 40 nm are used for the capping) and bismuth thickness of 5.51 $\mu$m. The spacing between the pixels further introduces a flat multiplicative factor in the instrument efficiency. State of the art fabrication processes allow reaching an average gap of 6.34 $\mu$m, leading to a 96 \% filling factor.

\begin{figure}[!h]
    \centering
    \includegraphics[width=16cm]{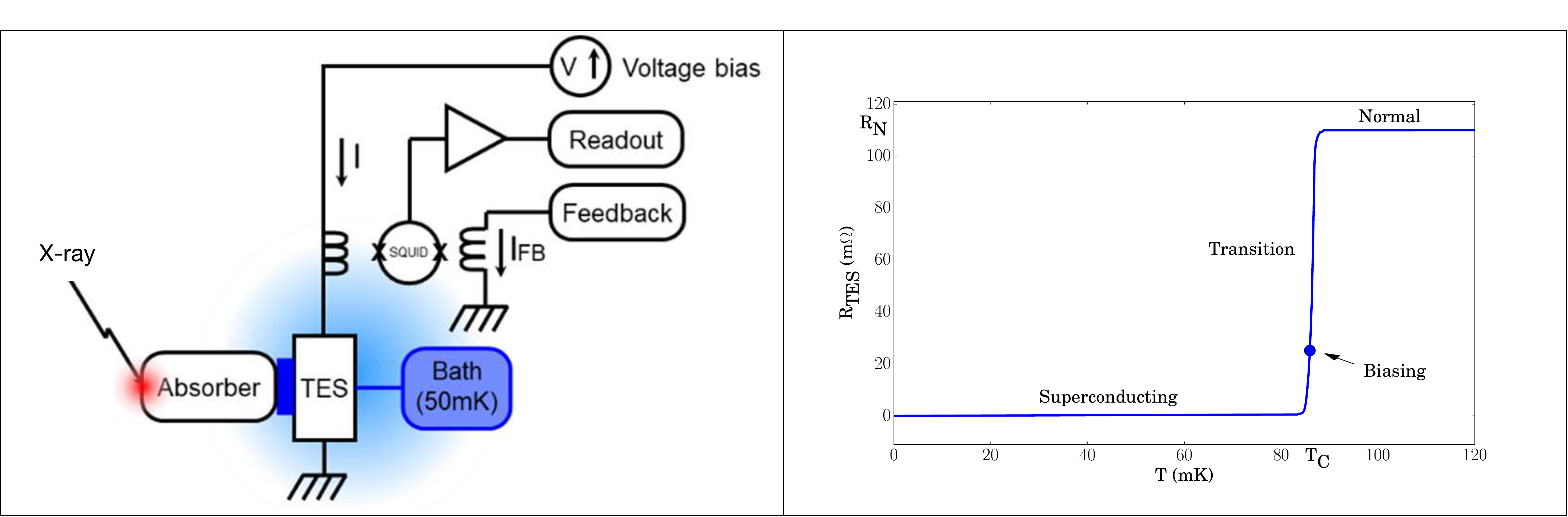}
    \caption{Left) The schematic of a TES. Right) The resistance variation as a function of the temperature, showing a sharp rise in the transition region between the superconducting and normal states\cite{Ravera_2014SPIE.9144E..2LR}.}
    \label{fig:tes_principle}
\end{figure}

The absorption of an X-ray in the absorber leads to a temperature increase of the TES/absorber couple and thus of the TES resistance (see Figure \ref{fig:tes_principle}). Under an (almost) constant bias voltage, the current going through the TES then shows a sharp decrease. Overall, each X-ray photon will create a current pulse, whose amplitude (and shape) depends on its energy. It is reconstructed on board to provide a precise measurement of the photon energy and arrival time. The combination of the TES/absorber heat capacity, the thermal conductance to the cold bath, and the electrical properties of the TES circuit, including the detector setpoint, define the temporal shape of the pulse. These are optimized to match the X-IFU 0.2 to 12 keV operating range and allow a multiplexed readout with minimal resolution degradation (see \ref{sec:CFEE}).\cite{Smith_2016SPIE.9905E..2HS,Smith_2021ITAS...3161918S}

\begin{figure}[!b]
    \centering
    \includegraphics[width=16cm]{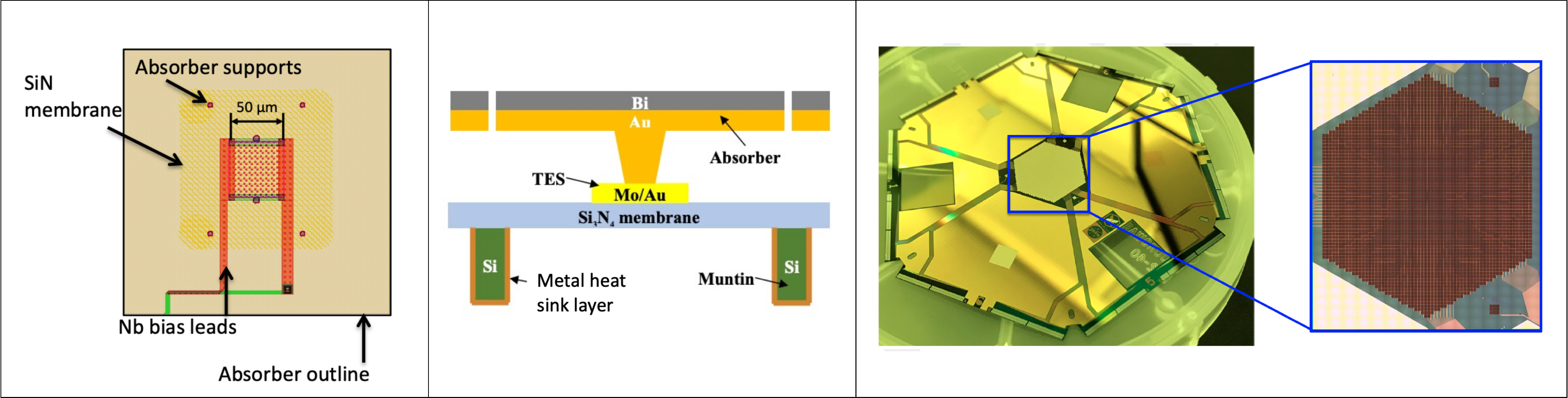}
    \caption{Left) Top-down view of pixel design. Center) Cross-section cartoon view of pixel layout. Right) Photograph of a X-IFU prototype hexagonal detector with more than 3000 pixels.}
    \label{fig:TES}
\end{figure}
\subsection{The cold front end electronics}
\label{sec:CFEE}
\begin{figure}
    \centering
    \includegraphics[width=17cm]{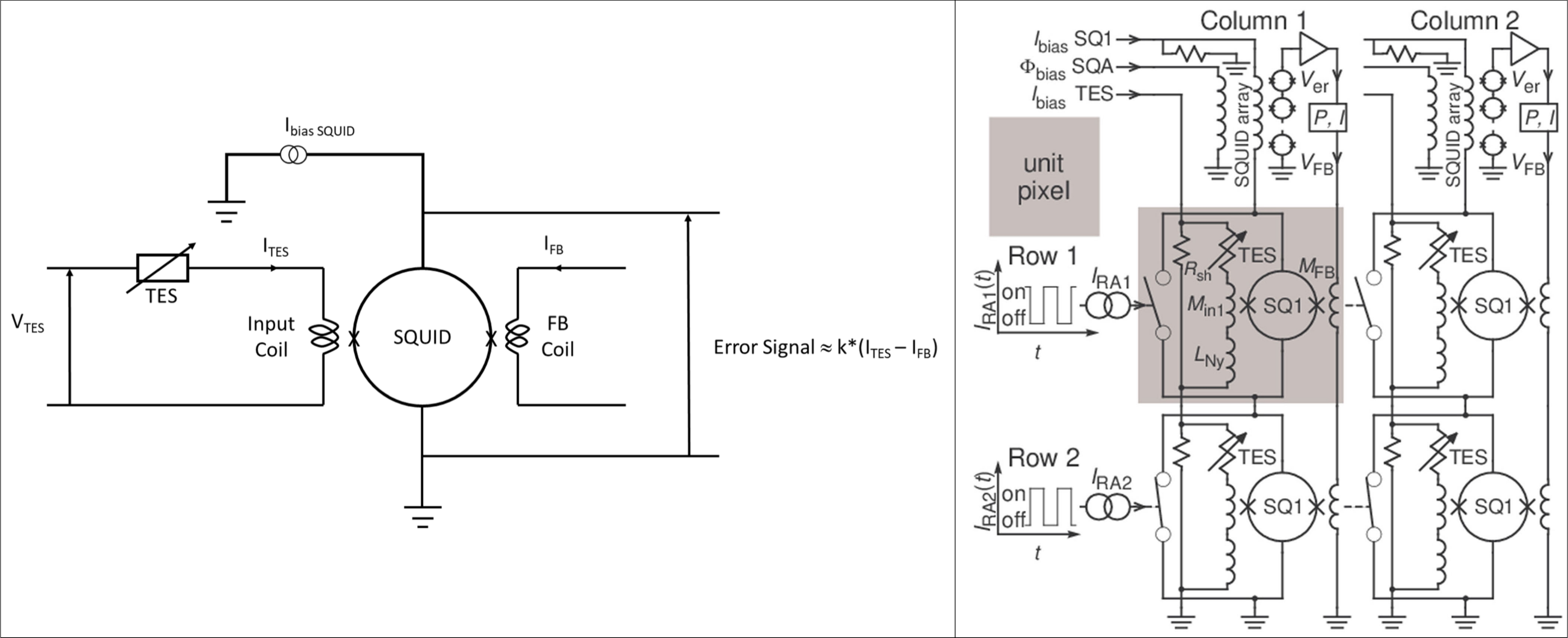}
    \caption{Left) Illustration of the feedback loop used to null out the magnetic field sensed by the SQUID\cite{Irwin_2005cpd..book...63I}. Right) Schematic\cite{Durkin_2019ITAS...2904472D} of 2-column × 2-row TDM. Each dc-biased TES is read out by a ﬁrst stage SQUID ampliﬁer (SQ1) via inductive coupling (M$_{in1}$). A row of SQ1s is turned on by applying a row address current (I$_{RA}$) to the corresponding row address line, opening the row’s ﬂux actuated switches. During TDM operation, rows are opened sequentially, reading out one TES per column at a time. Each column’s SQ1 signals are passed to a SQUID series array ampliﬁer, whose voltage (V$_{er}$) is read out by room temperature electronics.}
    \label{fig:feedback_loop_and_tdm_readout}
\end{figure}

The amplification of the signal generated by the TES relies on SQUIDs (Superconducting Quantum Interference Devices), split in two stages and referred to as MUX SQUIDs (also called SQ1) for the first stage, and as the AMP SQUIDs (sometimes named SSA for SQUID Series Array) for the second one\cite{Welty_1993ITAS....3.2605W,Kiviranta_2021ITAS...3160356K} respectively. SQUIDs are extremely sensitive magnetometers, operating at cryogenic temperatures. The current flowing through the TES is converted into a magnetic field by an input coil. To linearize the near-sinusoidal MUX SQUID response, and increase its dynamic range, the digital readout electronics generates a current for a feedback coil to null the magnetic field sensed by the SQUID (see Figure \ref{fig:feedback_loop_and_tdm_readout}). This error signal is then read out by the AMP SQUID, operated at 2K, to be amplified in order to increase the signal robustness against down-stream and external noise sources. The output voltage is then delivered to the low-noise amplifier of the WFEE operated at room temperature.

Because resources are limited on the spacecraft, multiplexing is required and the baseline adopted for X-IFU is Time Division Multiplexing (TDM\cite{Doriese_2016JLTP..184..389D,Doriese_2019ITAS...2905577D,Durkin_2021ITAS...3165279D}), see Figure \ref{fig:feedback_loop_and_tdm_readout} for a description of the TDM principles (an alternative readout technique would be frequency domain multiplexing\cite{Vaccaro_2022arXiv220812604V,Akamatsu_2020JLTP..199..737A,Akamatsu_2021ApPhL.119r2601A}). In TDM, each DC-biased TES is associated with a MUX SQUID. Those are activated sequentially via a ﬂux-actuated superconducting switch, so the TESs in each readout column are measured sequentially. Every SQ1 activation is called a row and a full set of samples of a column constitutes a readout frame. TDM columns are read out in parallel: the SQ1 of the same row number of each column are activated synchronously. The row addressing is performed in the Digital Readout Electronic (DRE). In order to limit the number of row address lines, two levels of switches are used: instead of having 34 pairs of command lines to command the ON/OFF state of the MUX SQUID, a command matrix of 9x4 is implemented, thus requiring only 13 pairs.  

In its present configuration, the total number of pixels of the detector array of X-IFU is 2376, divided in 6 groups of 396 pixels each. Each group of 396 pixels is read out by independent electronics, so that a failure would not lead to a loss larger than one sixth of the whole detector. One group of 396 pixels is thus made of 12 columns of 33(+1) pixels (33 rows with sensitive pixels $\times$ 12 columns = 396 pixels, plus one "dark row" with a pure resistor attached, that monitors the drift in the gain of the readout electronics). 
\subsection{The cryogenic anti-coincidence (\cryoac) detector and associated electronics}
\label{sec:cryoac}
\begin{figure}
    \centering
    \includegraphics[width=15cm]{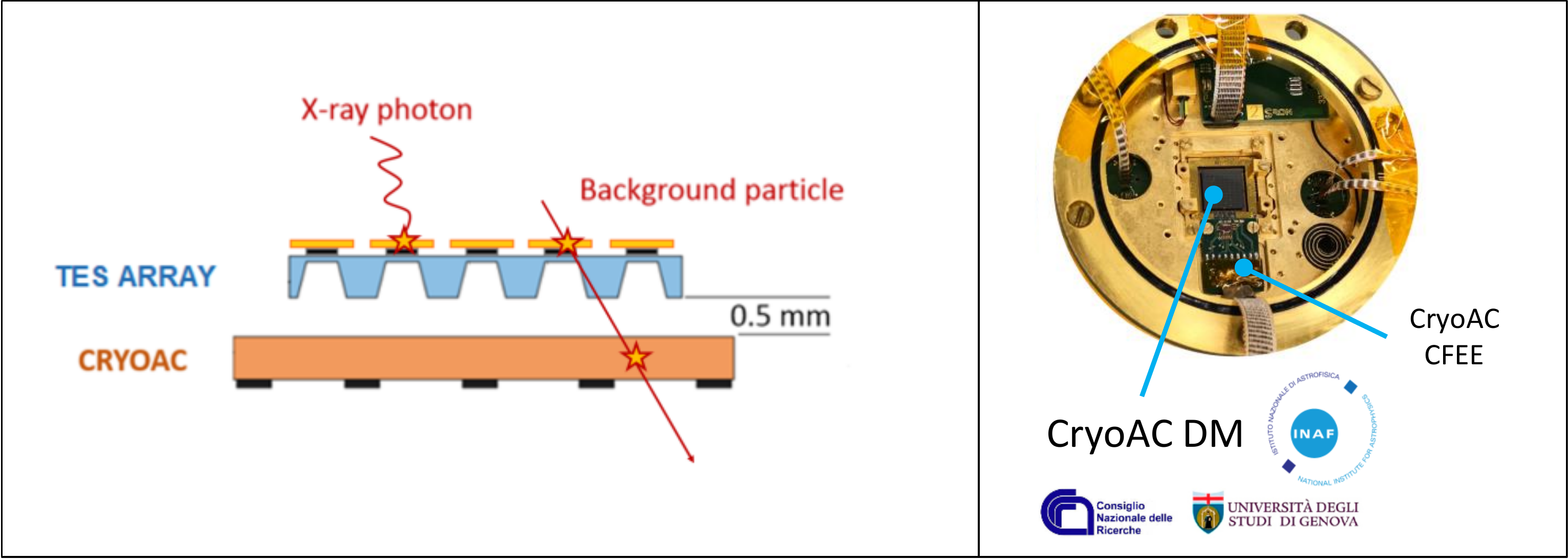}
    \caption{Left) Working principle of the \cryoac. Right) The \cryoac\ demonstration model and its cold front end electronics as used for the coupled test with prime micro-calorimeter array\cite{DAndrea_2022JLTP..tmp..139D}.}
    \label{fig:cryoac}
\end{figure}

Background will be generated by primary and secondary particles hitting the prime TES array and depositing energy in the science energy range of the X-IFU. A second cryogenic detector is accommodated just underneath the prime array (at a distance less than 1 mm, see Figure \ref{fig:cryoac}), so that events depositing energies both in the micro-calorimeter array and the \cryoac\ can be flagged out, and removed from the count stream from the observed source (see \cite{Lotti_2021ApJ...909..111L} for a recent review of the particle background of the X-IFU). The \cryoac\ is made of Si-suspended absorbers sensed by a network of IrAu TES\cite{Macculi_2016SPIE.9905E..2KM,Macculi_2020SPIE11444E..4AM,Macculi_2020JLTP..199..416M,DAndrea_2020JLTP..199...65D} (the first demonstration of the joint operation of the main TES array, and the \cryoac\ is presented here\cite{DAndrea_2022JLTP..tmp..139D}). Time coincident multiple events (generated primarily by secondary particles generated in the metallic parts close to the detector) can also be flagged out, although not necessarily having a time coincident event in the \cryoac. The \cryoac\ detector is segmented in four quadrants, each associated with its own readout electronics (\cryoac\ Warm Front End Electronics, \cryoac-WFEE and Warm Back End Electronics \cryoac- WBEE \cite{Macculi_2016SPIE.9905E..2KM,Macculi_2020SPIE11444E..4AM}). The \cryoac\ WFEE is divided in 4 quadrants, each one providing bias to TES and SQUID for a single pixel and producing the analog scientific signal and housekeeping. The \cryoac\ WBEE performs the digital processing, detecting, and time stamping cosmic ray events by applying a proper trigger logic algorithm (Chiarello et al., 2022, SPIE, submitted), and generating the associated telemetry. The WBEE function is implemented in two identical independent units operated in cold redundancy. Both units can operate the 4 anti-coincidence detection chains. Failure in one readout chain would mean that approximately one fourth of the prime array would have a higher background, though not leading to a complete loss as the pointing of the telescope could be offset so to avoid that particular quadrant. 
\subsection{The warm readout electronics}
Next we move to the two main components of the warm electronics of the main readout detection chain. 
\subsubsection{Warm Front-End Electronics (WFEE)}
\begin{figure}[!h]
    \centering
    \includegraphics[width=17cm]{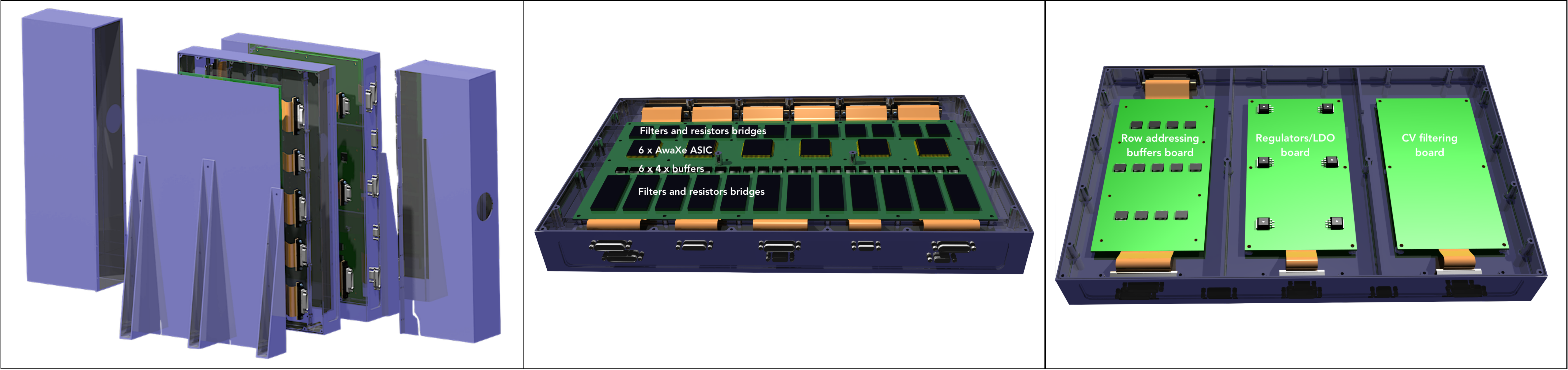}
    \caption{Exploded view of one WFEE box with its main components highlighted: the ASIC module with the buffers, the row addressing signal buffer module and finally the regulator module and the converter filtering board.}
    \label{fig:WFEE}
\end{figure}

The first stage of the warm electronics is called the Warm Front-End Electronics (WFEE\cite{Prele_020SPIE11444E..3UP}, see Figure \ref{fig:WFEE}). The main functions ensured by the WFEE are:
\begin{itemize}
  \setlength\itemsep{-2pt}
   \item  To amplify the error signal provided by the cold electronics (one by column),
    \item  To bias the TES (one TES bias per column),
    \item  To bias the 50mK and 2K SQUID,
    \item  To buffer the feedback signals,
    \item  To buffer the row addressing signal that controls the pixels selection.
\end{itemize}

As dealing with extremely sensitive signals, the WFEE must be accommodated as close as possible to the cryostat and share a Faraday cage with it to avoid noises generated by electromagnetic fields and common mode perturbations, and as far as possible from noisy digital electronics. The WFEE is implemented by 6 independent units, aligned on the 6 cold electronics groups. These 6 WFEE units are grouped in 3 electronic boxes. One unit thus corresponds to 12 readout chains.

\subsubsection{Digital Readout Electronics (DRE)}
\label{DRE}
\begin{figure}[!h]
    \centering
    \includegraphics[width=17cm]{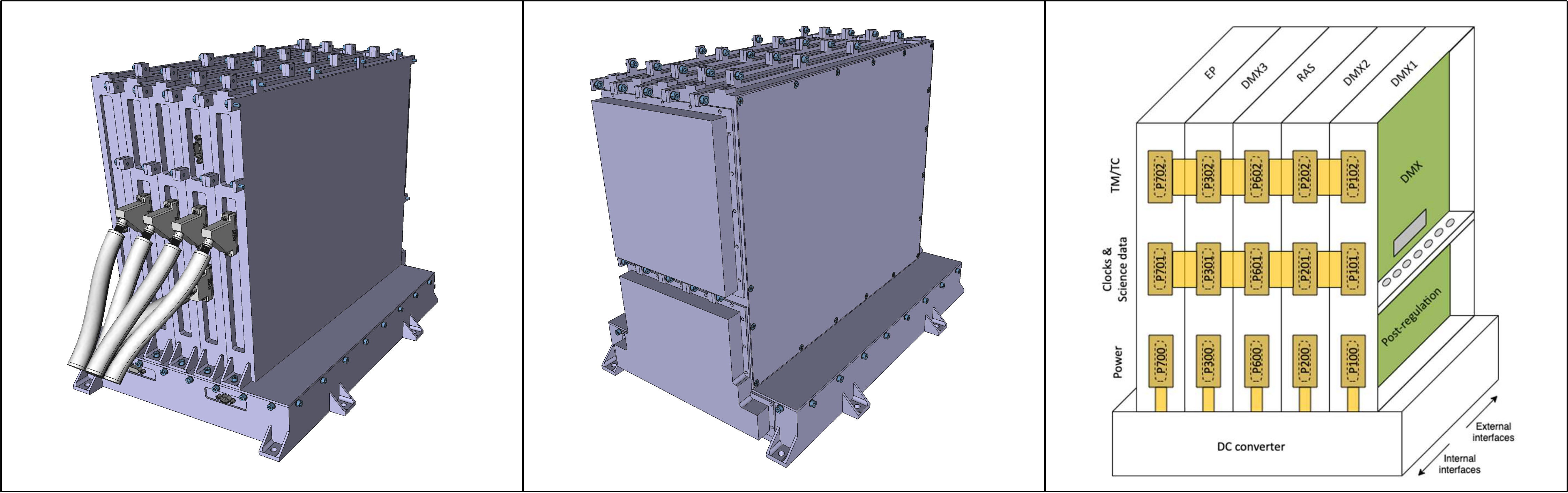}
    \caption{Left) Layout of a DRE unit, with the warm harnesses connecting to the WFEE. Center) The rear of a DRE unit, showing covers to protect internal interfaces against electromagnetic interferences. Right) Physical breakdown of a DRE unit, made of 3 demultiplexing (DEMUX) modules, 1 raw addressing and synchronisation (RAS) module and one module holding the event processor (EP). The DC converter is located at the bottom of the DRE Unit.}
    \label{fig:DRE}
\end{figure}

The second stage of the warm detection chain is the Digital Readout Electronics (DRE\cite{Ravera_2018SPIE10699E..4VR} for a description of the DRE prototype developed for Frequency Domain Multiplexing, but sharing some commonalities with the one being developed now for TDM readout, see Figure \ref{fig:DRE}). Its main functions are to:
\begin{itemize}
 \setlength\itemsep{-2pt}
 \item Generate the row addressing signals to select the row to be readout,
 \item Sample the error signal of the pixels of the selected row,
 \item Generate the feedback signal for the MUX SQUIDs,
 \item Generate dynamic offset compensation signals to the SSA to account for the offset dispersion of the different MUX SQUIDs output voltage inside a column
 \item Identify events in the pixel data flow and extract the data records (8192 samples each: the sampling frequency is 183.8 kHz equivalent to one sample every 5.44 $\mu$s),
 \item Process the data record to estimate the arrival time, to digitize the energy of the event and determine its grade\cite{Peille_2016SPIE.9905E..5WP,Cobo_2018SPIE10699E..4SC,Cobo_2020SPIE11444E..96C,Vega_2022PASP..134b4504V}. The accurate energy of the photon in keV will then be obtained on-ground accounting for the gain scale (and other corrections). 
 \item Produce telemetry packets and send them to the spacecraft mass memory.
\end{itemize}
The switching from one pixel to the other that results from the multiplexing generates high frequency transient signals along the detection chain that can propagate within one column and from one chain to its neighbors through parasitic coupling mechanisms. To minimize the impact of possible interferences, a synchronization is implemented so that the switching occurs at the very same time for the 72 chains, and the sampling of the signal can be done in a steady time window where no switching occurs. This synchronization is based on the distribution by the Instrument Control Unit (ICU) of 2 clocks to the 6 DRE (where the row addressing driving the switching is implemented), one at the frame frequency (183.8 kHz) and one at 62.5 MHz. The DRE FPGA master clock is derived from the latter. 



The DRE also implement an adjustable delay (several steps of 8 ns), that allows to correct static delays that may result from differences in harnesses length or electronics.

\subsection{The 2K Core}
\begin{figure}[!h]
    \centering
    \includegraphics[width=17cm]{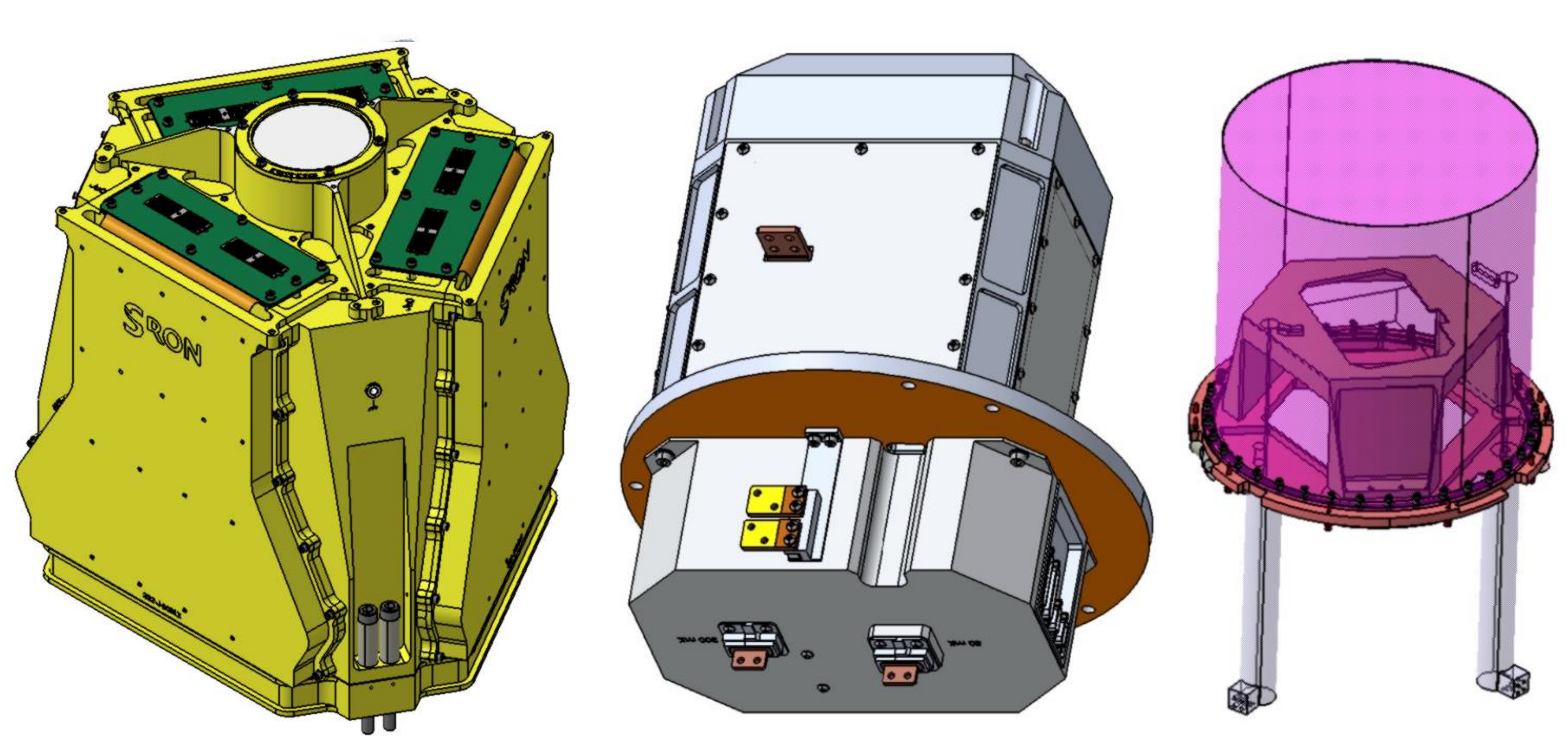}
    \caption{Left) The Focal Plane Assembly external layout, without FPA 2K baffle but with electrical interfaces with FPA Cold harnesses. Center) Hybrid Cooler Mechanical Assembly. Right) The 2K Frame}
    \label{fig:2KCore}
\end{figure}

The 2K Core consists of the Focal Plane Assembly (FPA) and the hybrid cooler cold mechanical assembly (CMA). It is placed inside the cryostat (see Figure \ref{fig:2KCore}). The so-called 2K Frame provides mechanical support to the FPA and CMA, and the thermal and mechanical interfaces with the cryostat. The 2K Frame provides also EMC tightness to the FPA, some harness routing, e.g. for the \cryoac\ signals, as well as thermal sensors and heaters. 

\subsubsection{Focal Plane Assembly (FPA)}
\begin{figure}
    \centering
    \includegraphics[width=17cm]{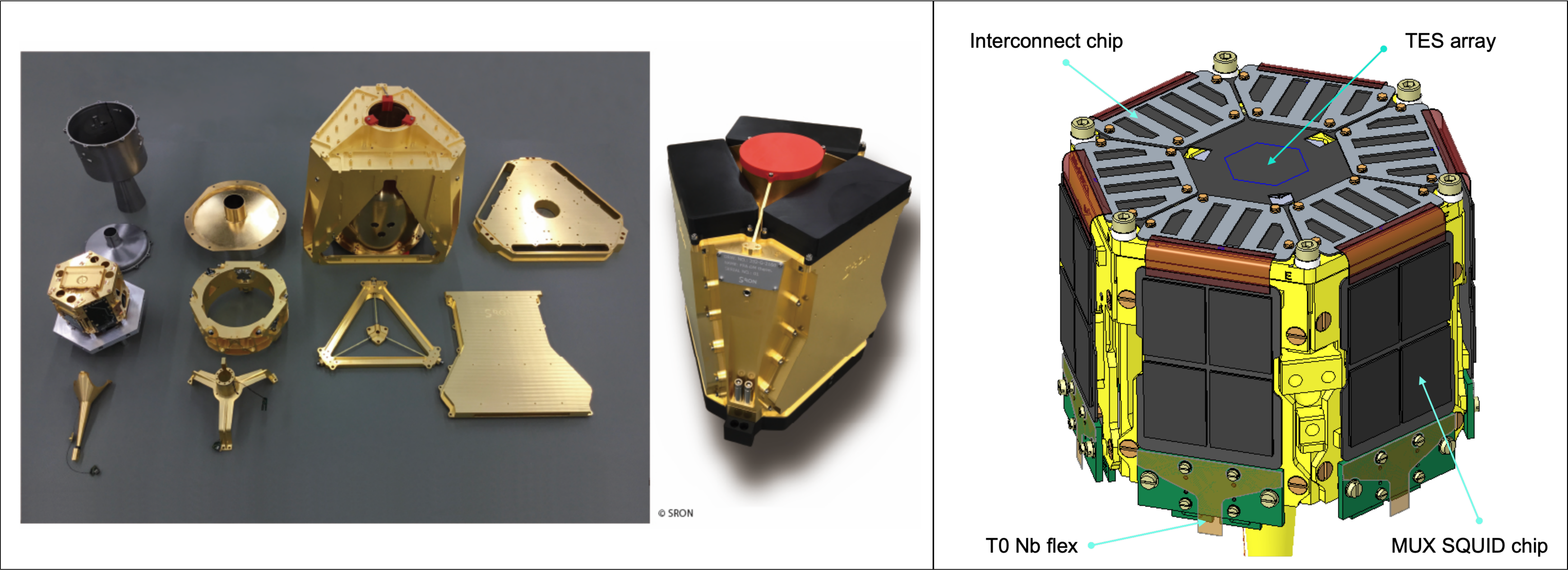}
    \caption{Left) Left: The various sub-assemblies of the FPA-DM during integration. At the top left the Niobium shield is shown. The triangle in centre is one of three Kevlar cord assemblies which, together, form the suspension of the T1 stage. The hexagonal structure on the left is the partially assembled detector stage (see right figure). On the right, the first FPA-DM completely assembled. The front-side red cap is a protective cover to avoid contamination of the optical filter. The three front black covers will be replaced by the readout harness during further tests. The black back-side cover protects the sensitive electronics and internal harness of the FPA during transport and handling (van Weers et al., 2022, SPIE, submitteed and AthenaNuggets43). Right) Preliminary design concept of the T0 (50 mK) detector assembly. The TES array is mounted at the top of the structure and wire-bounded to an interconnect chip. Each carrier/interconnect chip holds 4 MUX SQUID chips with Indium bump-bonds realizing the required high-density, low-impedance contacts between the carrier and MUX SQUID chips. As a baseline, the Nb flex harness from the FPA's T2 stage will be glued to a rigid support that is mounted below the Carrier chip, allowing wire-bonding directly from the Nb flex harness to the carrier chip.}
    \label{fig:FPA}
\end{figure}

The FPA\cite{Jackson_2016SPIE.9905E..2IJ} hosts the two cryogenic detectors (prime micro-calorimeter array and \cryoac) and their cold readout electronics, further providing the thermal isolation and electromagnetic shielding required to operate the sensor array at $\sim 50$ mK within the environment of the 2K Core, located inside the instrument cryostat. An exploded view of the FPA is shown in Figure \ref{fig:FPA} (see van Weers, 2022, SPIE, submitted).

The FPA implements the following functions: 
\begin{itemize}
 \setlength\itemsep{-2pt}
\item the main TES detector array operated at nominal temperature T0 stage (50 mK), 
\item the cryogenic input electronics: 1) MUX SQUIDs for the main array at nominal temperature T0 stage; 2) AMP SQUID at nominal temperature T2 stage (2 K),
\item the magnetic and electric field shielding from the environment,
\item optical filtering of low-energy out-of-band photons from both sky and on-board metrology sources, 
\item the \cryoac\ detector. 
\end{itemize}

\subsubsection{The hybrid cooler}
\begin{figure}
    \centering
    \includegraphics[width=17cm]{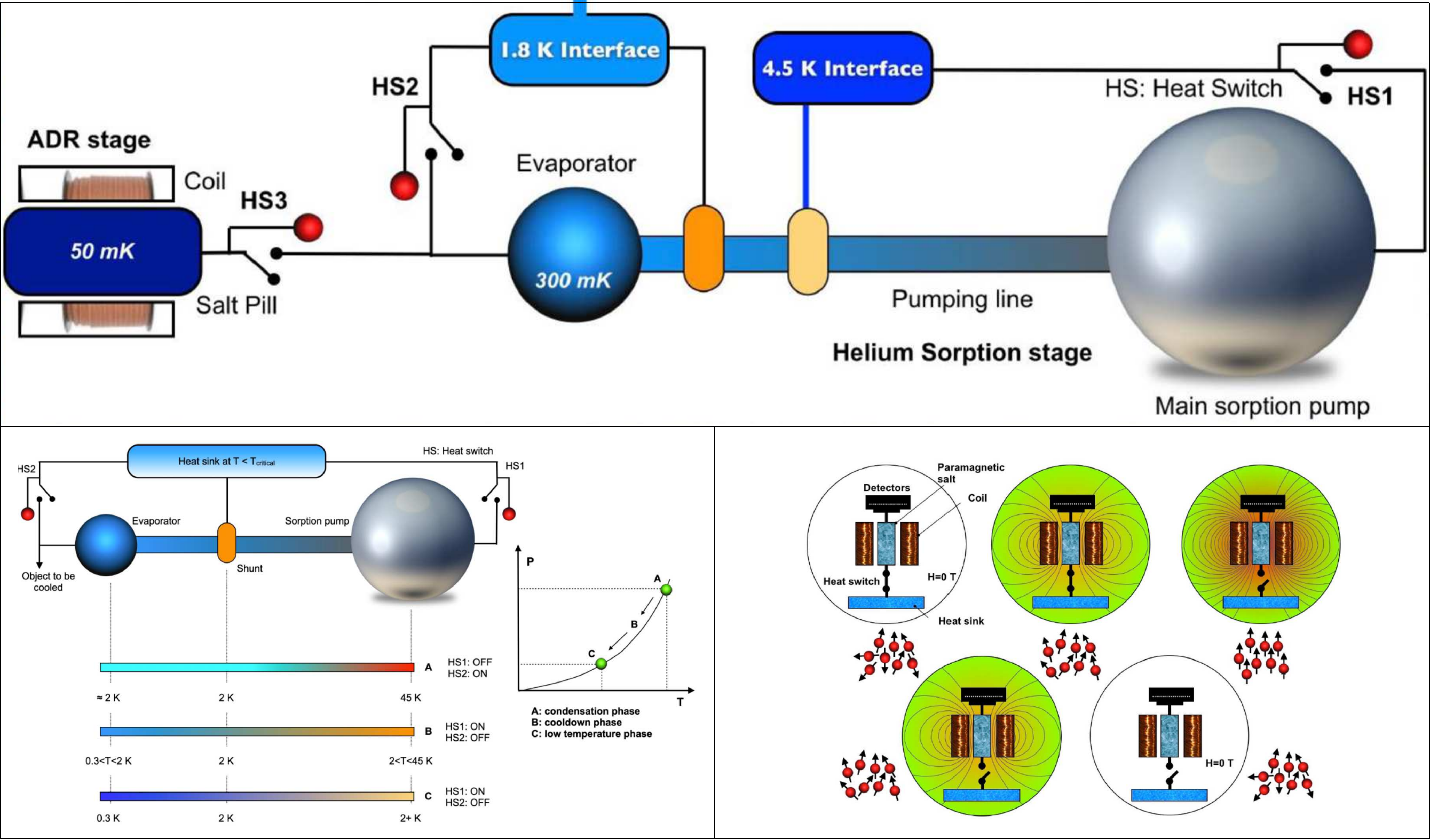}
    \caption{Top) The schematic of the Hybrid cooler mechanical assembly. Bottom left) The working principle of the sorption stage. Bottom right) The working principle of an Adiabatic Demagnetization Refrigerator (ADR).}
    \label{fig:hybrid_cooler}
\end{figure}

The hybrid cooler implements the following thermal functions: 
\begin{itemize}
 \setlength\itemsep{-2pt}
    \item To provide and control cooling power at a T0 thermal interface (50 mK range): this function provides cooling power to the detector stage of the FPA, 
    \item To provide cooling power at a T1 thermal interface (300 mK range): this function provides cooling power to the T1 stage of the FPA (intercept of heat loads from T2 stage to T0 stage),
    \item To reject heat at T2 (1.9 K range) and T3 (4.9 K range): this function allows for the evacuation of the heat generated by the sub-Kelvin cooling on the two heat sinks provided by the upper cooling chain.
\end{itemize}
 The Hybrid cooler\cite{Duband_2014SPIE.9144E..5WD,Duband_2017SPIE10566E..0AD} combines 1) a $^3$He sorption cooler, providing the cooling power at T1 (300 mK range), and 2) an Adiabatic Demagnetization Refrigerator (ADR), providing the cooling power at T0 (50 mK range). The total cooling power is $\sim 1 \mu$W and $\sim 15-20\mu$W at 50 and 300 mK, respectively (see Figure \ref{fig:hybrid_cooler}), providing significant margins with respect to the currently estimated  FPA dissipation (of order 500 nW and 6-7$\mu$W respectively).

The sorption stage features: 
\begin{itemize}
 \setlength\itemsep{-2pt}
    \item a sorption pump,
    \item an evaporator, into which liquid $^3$He is condensed and then pumped down, with a thermal link to the objects to be cooled: the ADR stage of the cooler, and the T1 stage ($\sim 300$ mK) of the FPA,
    \item  a pumping line, between the pump and the evaporator,
    \item thermal links toward heat sinks at T2 (1.9K) and T3 (4.9 K), provided by the upper cryochain,
    \item two gas gap heat switches (HS1 and HS2), allowing to thermally connect or disconnect respectively the sorption pump to the T3 interface, and the evaporator cold end to the T2 interface.
\end{itemize}
This system is a one-shot system and needs to be recycled once all the liquid helium has been exhausted. During the recycling phase, the sorption pump first is heated to exhaust most of the helium gas and allows good condensation efficiency. Then once the liquid has been condensed, the sorption pump is cooled back down to ensure the pumping effect and lower the saturated vapor pressure of the liquid helium bath.

The ADR stage is based on magnetic cooling, which principle is based on the entropy reduction of the disordered spins of paramagnetic material: 1) when submitted to a magnetic field, the spins in the paramagnetic material tend to align along the field direction; the order in the material increases, thus the entropy decreases, and the material releases heat toward a heat sink; 2) when the magnetic field is gradually reduced and turned off, the spins turn back to their random orientation, which requires energy; if the material is thermally isolated, this energy is drawn from the thermal agitation, which cools the material in an adiabatic process. Thus, like the sorption stage, the ADR stage has a cyclic operation, with a cooling phase followed by a recycling phase.
The ADR stage is composed of: 
\begin{itemize}
 \setlength\itemsep{-2pt}
    \item a cryogenic superconductor coil,
    \item a paramagnetic salt pill accommodated inside of the coil,
    \item a thermal link to the object to be cooled: the T0 stage ($\sim 50$ mK) of the FPA (detector stage),
    \item a third heat switch (HS3), allowing to thermally connect the salt pill to the cold end of the sorption stage evaporator during the recycling phase, to thermally disconnect the salt pill from the rest of the cooler during the cooling phase.
\end{itemize}

The Hybrid Cooler will provide a minimum cold time of 28.5 hours, while the total recycling time should not exceed 8 hours and 50 minutes. A partial regeneration of the cooler is also possible. 

The CMA is controlled by 2 independent warm electronics, operated in cold redundancy, that enable the control of the cooler (e.g., cold tips temperature regulation, regeneration) as well as any telemetry needed to operate the cooler.
\subsection{The thermal filters}
\label{sec:filters}
\begin{figure}
    \centering
    \includegraphics[width=17cm]{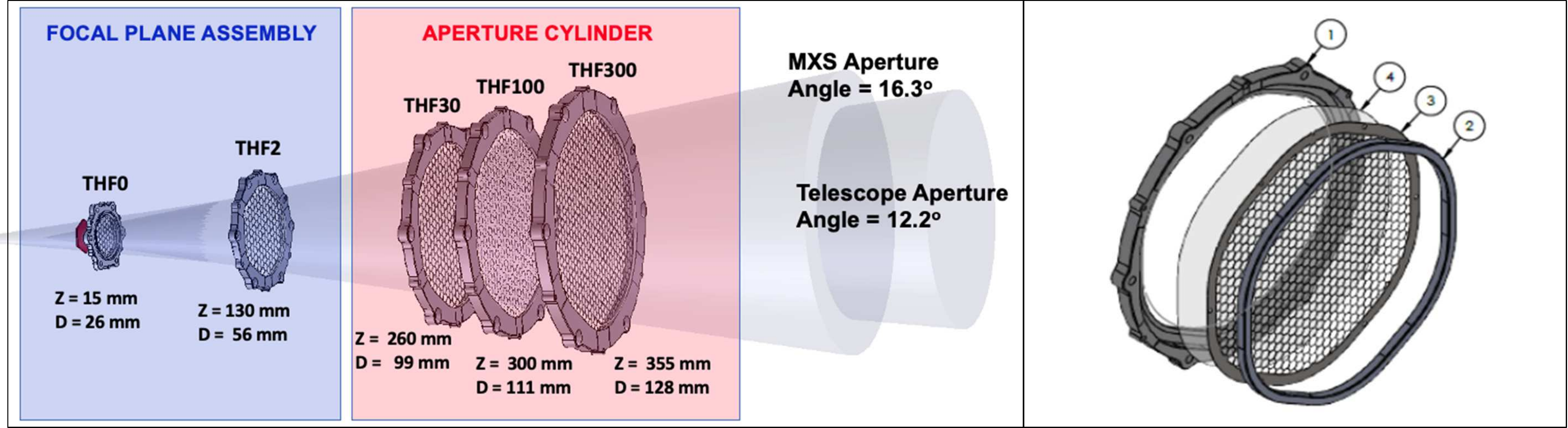}
    \caption{Left) The stack of thermal filters located in the X-IFU field of view. Three filters have interfaces with the cryostat through the aperture cylinder, and 2 filters (the coldest) are located in the FPA. Right) THF300 schematic mounting of the film and supporting mesh inside the two-parts type frame. \textcircled{1} is the outer frame, \textcircled{2} is the inner frame, \textcircled{3} is the supporting mesh and \textcircled{4} is the thin polyimide/Al foil. }
    \label{fig:thermal_filters}
\end{figure}

The thermal filter (THF) stack is a set of 5 filters accommodated in the field of view of the X-IFU\cite{Barbera_2014SPIE.9144E..5UB,Barbera_2016JLTP..184..706B,Sciortino_2016SPIE.9905E..66S,Barbera_2018JLTP..193..793B,Barbera_2018SPIE10699E..1RB,Locicero_2018SPIE10699E..4RL,Sciortino_2018SPIE10699E..50S,Sciortino_2018JLTP..193..799S,Puccio_2020JATIS...6c8003P}. The functions implemented by the thermal filters are:
\begin{itemize}
 \setlength\itemsep{-2pt}
    \item To protect the detector from photons shot noise and thermal radiations,
    \item To ensure the EMC tightness of the 2K Faraday cage,
    \item To ensure the EMC tightness of the 300K Faraday cage.
\end{itemize}
These functions have to comply to the following set of constraints: 1) transparency to X-rays, 2) temperature interfaces set by the cryostat and the FPA, 3) thermal loads onto colder stages, 4) mechanical loads from launch vibrations, 5) differential pressure during instrument pumping, 6) Thermal conductivity across the filter opening. Figure \ref{fig:thermal_filters} shows a schematic diagram of the THF stack with specified distances from the focal plane (Z, in mm), clear aperture diameter (D, in mm). Filters are named by the approximate temperature of the shield they are mounted on.

In its current design, each filter is composed of 30 nm of Aluminum (of which 7 nm are considered to be oxidized to Al$_2$O$_3$) and 45 nm of polyimide. The meshes of all filters were optimized for a blocking factor between 2 and 2.5\%, the colder (and thus smaller) filters having a poorer transmission due to the minimal thickness of the mesh wires. This leads to a total loss of $\sim 10.7 $\%.

\subsection{The Aperture Cylinder (ApC)}
\begin{figure}[!h]
    \centering
    \includegraphics[width=17cm]{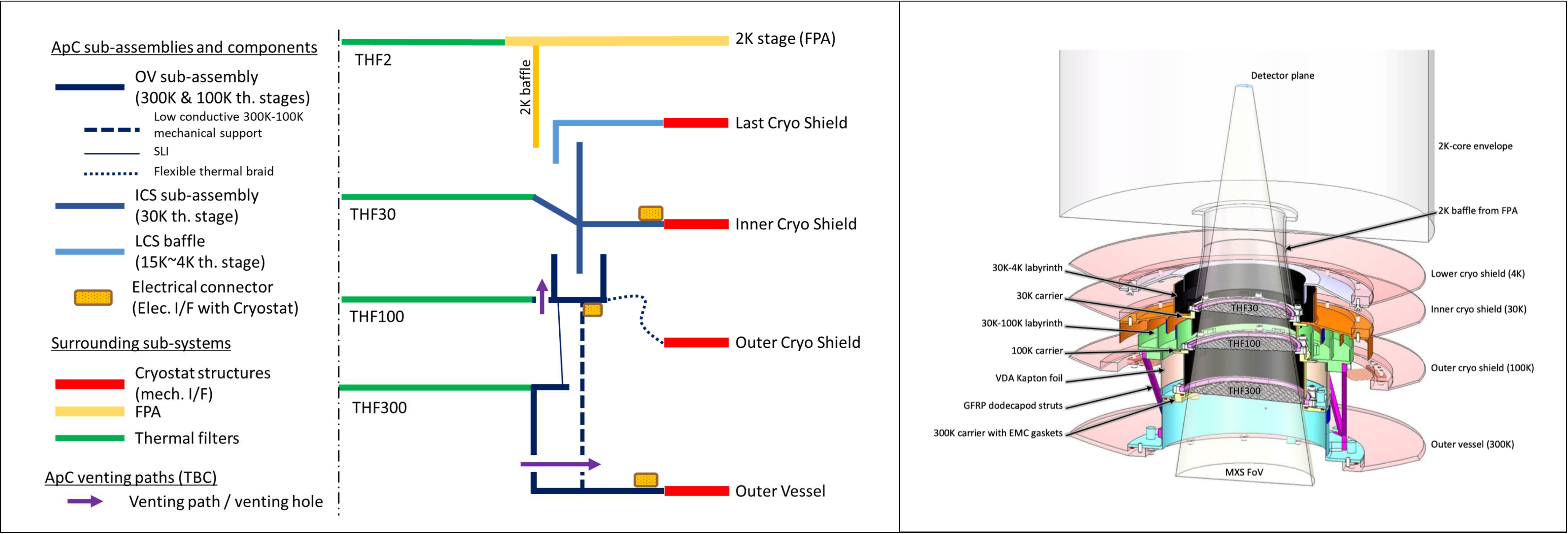}
    \caption{Left) The ApC architecture and interfaces. Right) Drawing of the mechanical design of the ApC. See Thibert et al., 2022, SPIE submitted.}
    \label{fig:apc}
\end{figure}

The Aperture Cylinder (ApC), together with the stack of thermal filters, forms a feed-through assembly that allows X-rays to reach the TES array, with minimal absorption in the useful energy range 0.2–12 keV (see Figure \ref{fig:apc}).
The ApC contributes to, or ensures, the following functions of the X-IFU:
\begin{itemize}
 \setlength\itemsep{-2pt}
    \item To ensure the EMC tightness of the 300K Faraday cage,
    \item To mechanically support the filters,
    \item To ensure the thermalization of the filters with respect to the cryostat interface temperature,
    \item To protect the detector from photons shot noise and thermal radiations,
    \item To protect the filters from Dewar contamination,
    \item To provide heating capability to the thermal filters,
    \item To reduce straylight radiation to the detector,
    \item To protect the detector against the radiated electromagnetic fields.
\end{itemize}

\subsection{The Filter Wheel (FW)}
The Filter Wheel used for X-IFU implements the following functions\cite{Bozzo_2016arXiv160903776B}: 
\begin{itemize}
  \setlength\itemsep{-2pt}

    \item Reduction of the optical load from bright stars which could degrade the energy resolution of the instrument,
    \item Possibility to optimize the observational throughput in case of bright X-ray targets,
    \item Provision of a radioactive calibration source ($^{55}$Fe) as a backup to the Modulated X-ray Sources (MXS) and for early calibration purposes,
    \item Capability to monitor the intrinsic instrumental background,
    \item Powering and controlling the Modulated X-ray Sources (flash duration, intensity, frequency),
   \item Protection of the X-IFU focal plane detectors against micro meteorites and contamination.
   \end{itemize}
\begin{figure}[!h]
    \centering
    \includegraphics[width=17cm]{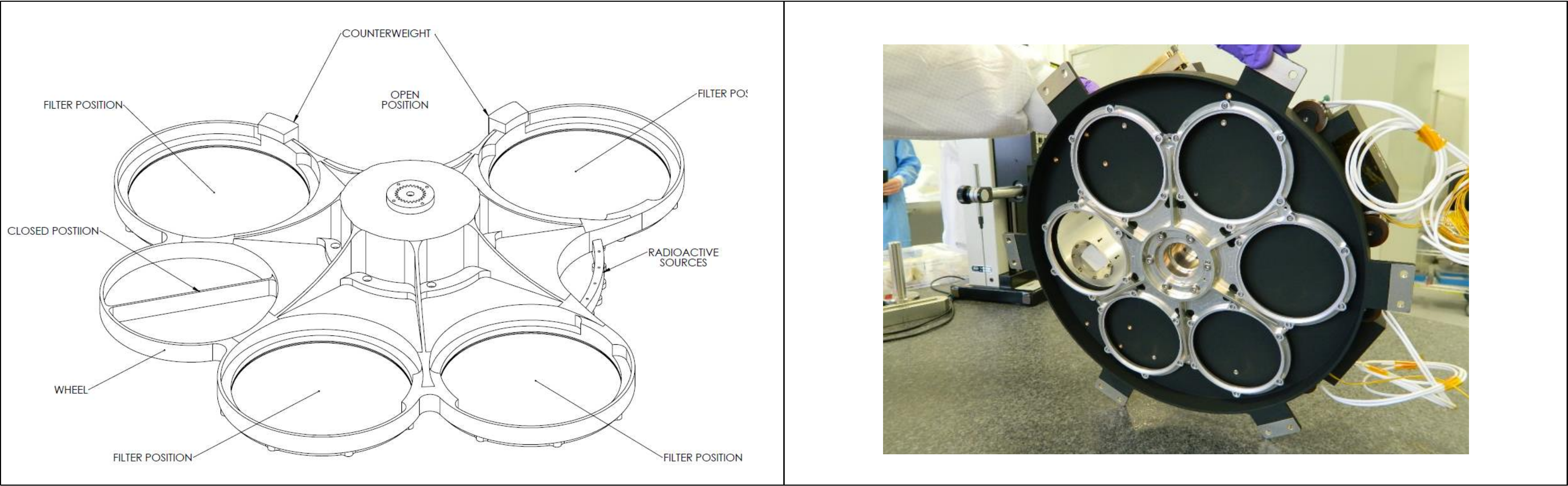}
    \caption{Left) The various filter positions of the X-IFU FW. Right) The FW developed for Hitomi by the University of Geneva and the Swiss industry Ruag Space. The approximate size of the overall FW shown above is 351 mm (length) $\times$ 384 mm (width) $\times$ 100 mm (height)\cite{Bozzo_2016arXiv160903776B}. The X-IFU FW having one more filter will be significantly larger in diameter ($\sim 68$ cm). }
    \label{fig:filterwheel}
\end{figure}
The current implementation of the filter wheel foresees (see Figure \ref{fig:filterwheel}): 
\begin{itemize}
 \setlength\itemsep{-2pt}
    \item One closed position (Molybdenum filter), to be used during launch for the protection of the focal plane detectors and in-flight for the evaluation of the internal detectors background.
    \item One open position (no filter) that is used for most observations or in the event of failures related to the FW.
    \item One thick filter with a transmission equivalent to 100-$\mu$m of Beryllium  to suppress the X-ray fluxes of celestial sources at energies $< 3$ keV. This filter can thus be adopted to limit the degradation of the X-IFU performances during the observations of particularly soft and bright X-ray sources.
    \item One thinner filter with a transmission equivalent to 20-$\mu$m of Beryllium for the observation of X-ray sources of intermediate brightness (while preserving more soft X-ray photons).
    \item Two optical blocking filters that are needed to limit the optical load (and thus the degradation of the instrument energy resolution) from the bright UV/V counterparts of the X-ray sources to be observed by the X-IFU. The first filter will be made out of a 200 nm-thick polyimide foil covered with a layer of 30 nm-thick Aluminum layer. The second filter will be made out of a 200 nm-thick polyimide foil covered by a 60 nm-thick Aluminum layer.
    \item A number of $^{55}$Fe radioactive sources, to be used for the calibration of the X-IFU.
\end{itemize}

The filter wheel mechanical assembly (FWA) is controlled by its associated electronic box: the filter wheel electronics (FWE).
\subsection{The Calibration Assembly (CAS)}
\label{sec:cas}
To capitalize on the good intrinsic energy resolution of the X-IFU main detector, it is critical to be able to correct for drifts in the detector response and in the electronics. These drifts can be caused by many factors including the thermal load on the detector, its magnetic field and temperature variations in the electronics, etc... They typically need to be corrected to a level of better than a few 0.1 eV. The drifts are assumed to be slowly varying and being able to correct these on a 4000 s time scale is sufficient. 

X-ray emission lines, if intense enough, allow such post-facto drift corrections. Natural X-ray sources (such as a $^{55}$Fe source) would be bright enough but continuously illuminate the detector (can be placed on a position in the filter wheel) but would contribute to the instrumental background if used during the observation. Therefore, modulated X-ray sources (MXS), similar to the one developed for Hitomi, are the natural choice\cite{deVries_2018SPIE10699E..65D}.
\begin{figure}[!h]
    \centering
    \includegraphics[width=17cm]{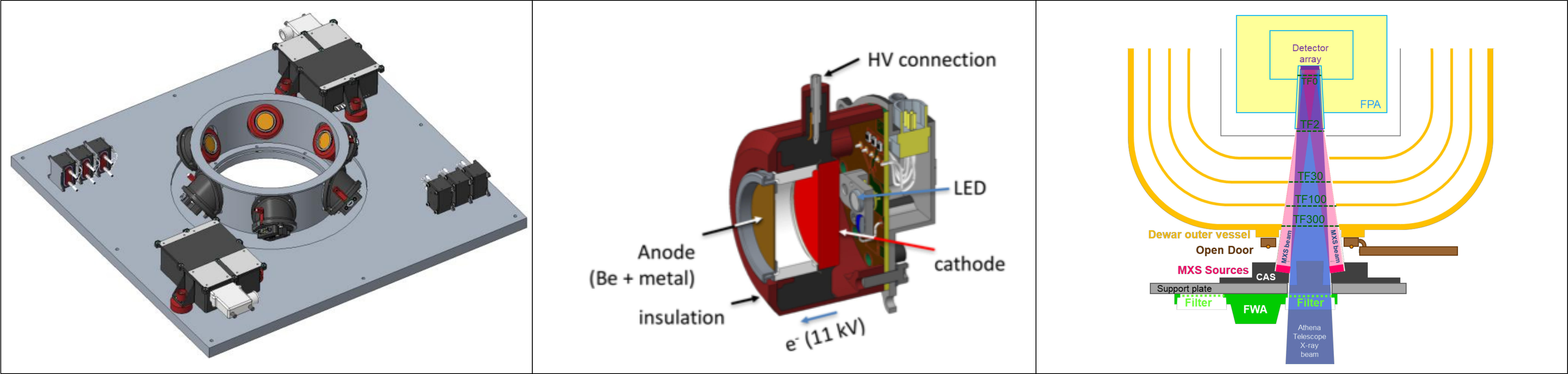}
    \caption{Left) View of CAS with 6 MXS and 2 High Voltage Power Supply units (I-PRR Design). Center) One MXS Right) FW and CAS accommodation wrt Athena telescope X-Ray beam and detector visibility cone.}
    \label{fig:cas}
\end{figure}
The CAS includes  (see Figure \ref{fig:cas}) : 
\begin{itemize}
 \setlength\itemsep{-2pt}
    \item six Modulated X-ray Sources (MXS),
    \item a CAS ring carrying the MXS around the telescope X-ray beam,
    \item the high voltage harness linking the HVPS to the MXS,
    \item two High Voltage Power Supplies (HVPS),
    \item a CAS support plate.
\end{itemize}

 The CAS, alongside the FW, is mechanically supported by the Filter Wheel support Structure (FWSS, plate), to be accommodated in the SIM at the entrance of the instrument (see Figure \ref{fig:cas}). The MXS photon flashes will be synchronized with the instrument time reference, thereby allowing their identification in the science telemetry. The flashes will typically last 1 second and be generated every 67 seconds. 

As stated above, the MXS are controlled by the FW electronics.

\subsection{The 2K Core Controller}
The 2K Core Controller (2KCC) is a unit which groups all the items controlling the 2K Core in a single box. Its main function is to perform the internal thermal control of the 2K Core, and in particular of the T0 (50 mK) stage of the FPA. The 2KCC merges the Hybrid Cooler Controller electronics (HybCCE) and the FPA Auxiliary Board (FAB).

The HybCCE is in charge of controlling the hybrid cooler. Its functions are:
\begin{itemize}
 \setlength\itemsep{-2pt}
    \item To provide electrical power to the ADR magnetic coil and the associated $^3$He sorption stage,
    \item To perform  temperature and voltage acquisitions of the different items in the cooler,
    \item To manage the temperature regulation of the cooler, based on the cooler temperatures and ADR current control.
\end{itemize}
The FAB is in charge of the: 
\begin{itemize}
  \setlength\itemsep{-2pt}
   \item Housekeeping of the FPA (readout of FPA thermal sensors),
    \item Powering the FPA heaters,
    \item Control the FPA B-coil used to optimize the magnetic field at the detector interface.
\end{itemize}

Additionally, the 2KCC is in charge of the housekeeping (thermal sensors readout and heaters activation) for the 2K Core frame (2KC HK) (2K plate, 2K EMC covers, 2K and 4K thermal links, etc.). The 2KCC also includes a DC/DC converter (also called LVPS or CV) for the supply of secondary power to the other parts of the unit (FAB, HybCCE \& 2KC HK).

\subsection{Cold harnesses}
The signals generated within the cryostat must be propagated to the warm electronics. This is achieved via cold harnesses, which are divided in 5 bundles: 1) 3 harnesses for the FPA Detector signals, 2) 1 harness for the \cryoac\ signals and including the 2K core housekeeping (HK) and control signals, and 3) 1 harness for the Hybrid Cooler ADR Coil current lead. Those harnesses are a critical part of any cryogenic system, as they must fulfill several requirements: 1) low thermal loads, implying the use of ultra thin, low thermal conductance wires (stainless steel or phosphor bronze, as thin as 100$\mu$m), 2) a careful optimization between joule dissipation and conductive losses, 3) a limited crosstalk between wires, and 4) a large number of wires to be accommodated via connectors on the cold side, where room is highly constrained.  

\begin{figure}[!th]
    \centering
    \includegraphics[width=17cm]{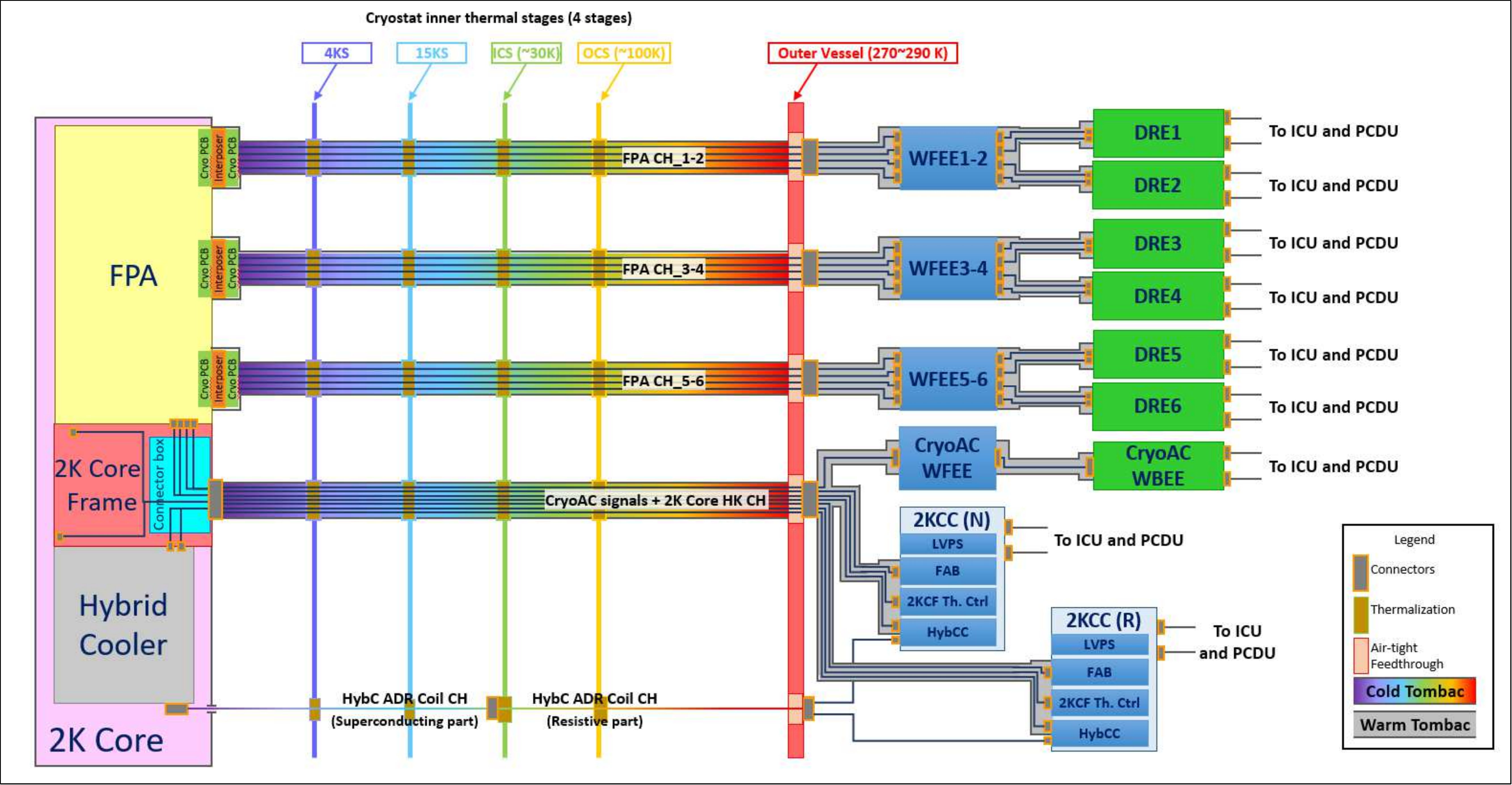}
    \caption{Cold harness overview, showing the three types of harnesses used for X-IFU. }
    \label{fig:cold_harnesses}
\end{figure}

There are two technologies to be considered for X-IFU: Twisted Shielded Pairs (TSPs) versus looms. The current understanding is that TSP wires are better for performance, but more delicate for manufacturing and implementation, and more demanding or constraining in terms of interfaces (rejected heat loads, mass, volume, stiffness, bending radius). Looms are easier to implement and less constraining, but may have a worse performance. Because it is believed that manufacturing and thermal issues can be overcome, TSPs are currently the preferred option. The wiring of the cold harnesses for X-IFU is presented in Figure \ref{fig:cold_harnesses}.
\subsection{The Instrument Control Unit}
The ICU embeds the instrument central software (Boot software and Application software) and provides all the communication and synchronization capabilities to command and control the instrument. It is the unique communication node with the satellite: in flight, all the commands and telemetry are normally exchanged through the ICU. In order to keep the ICU as independent from other units as possible, the instrument design limits the complexity of the interfaces enabling the ICU control and command: 1) All instrument's units are connected to the ICU through an internal SpaceWire (SPW) network. The ICU is connected to the spacecraft through a SPW link.

The Instrument Control Unit (ICU) implements the following functions:
\begin{itemize}
 \setlength\itemsep{-2pt}
 
 \item To synchronize photon readout. This function enables a synchronous readout of the pixels so that transient perturbations generated by the multiplexing (when switching from a pixel to another) do not happen when the sampling of other readout chains occurs (see \S \ref{DRE})
 
 \item To manage instrument internal communications. This function provides the X-IFU units with a low level bi-directional capability of exchanging digital information 
 \item To manage platform communications. This functions provides the X-IFU with a capability to send standardized space packets to the spacecraft, and to receive standardized space packets from the spacecraft 
 \item To manage observability and commandability
 \item To distribute time information to subsystems. This function distributes time information (time references, time broadcast…) from which units needing a time stamping capability can derive an adequate time reference
 \item To ensure instrument safety. This function monitors critical instrument parameters, and performs (or requests) the passivation actions (including from the spacecraft) needed to avoid failure propagation inside the instrument
 \item To manage instrument expertise. This function gathers all the observability and commandability services needed to monitor the performance of the main readout chain. As the needs related to expertise go far beyond the classical command and control function, it is presented as a separated function
 \item To manage instrument operability (modes transitions, autonomy…). This function gathers the highest level of autonomy services needed to implement the instrument operations: recycling sequences, detectors setup, calibration sequences…
\end{itemize}

It is made of a single box, hosting two independent units used in an adapted cold redundancy concept: only one unit is used as the instrument controller, but the second one could be switched ON for maintenance (mainly software upload or troubleshooting). In this case, the first unit controls the instrument, and the second is not allowed to interact with other instrument units.

The spacecraft and instrument units are connected to a 100 Mbps Space Wire router implemented in the ICU. This architecture enables high date rate communications for instrument management purposes and science telemetry transfer.
\section{Mass and Electrical Power budgets}
The budgets presented below are the current  estimates at the time of the X-IFU SRR. They would still remain subject to changes in later phases of the project.   
\subsection{Mass budget}
The X-IFU mass budget is given in Table \ref{tab:mass_budget}. This does not include the cryostat and the cooling chain, which are not in the X-IFU Consortium perimeter. The not-to-exceed mass allocation for X-IFU is 270 kg. 
\begin{table}[!h]
    \centering
    \begin{tabular}{|l|c|}
    \hline
    Unit & Mass (kg) \\
    \hline
        Hybrid Cooler Mechanical Assembly & 8.6 \\
        Focal Plane Assembly & 11.3 \\
        Connector boxes & 2.2 \\
        2K Core frame & 8.1 \\
        sub-total 2K Core & 30.2 \\
        \hline
        Aperture Cylinder & 4.3 \\
        \hline
        Filter Wheel & 13.0 \\
        Filter Wheel Supporting Structure & 8.6 \\
        Calibration assembly & 5.3 \\
        sub-total & 26.9 \\
        \hline
        Warm Front End Electronics & 24.9 \\
        Digital Readout electronics & 102.4 \\
        \cryoac\ Warm Front End Electronics & 2 \\
        \cryoac\ Warm Back End Electronics & 4 \\
        Instrument Control Unit & 15 \\
        Filter wheel electronics & 3.2 \\
        2K Core Controller & 22 \\
        sub-total e-Boxes & 173.4 \\
        \hline
        FPA Cold Harness & 15 \\
        2K C-HK and CryoAC Cold Harness & 5 \\
        Hybrid cooler coil harness & 1 \\
        sub-total cold harnesses & 21 \\
        \hline
        Total X-IFU & 221.3 kg \\
\hline
     \end{tabular}
    \caption{The nominal mass budget of X-IFU to be compared with the 270 kg not-to exceed allocations.}
    \label{tab:mass_budget}
\end{table}
\subsection{Electrical power Budget}
The electrical power budgets are provided in Table \ref{tab:electrical_power_budget} for  three sizing instrument modes, excluding again the cryogenic chain: 
\begin{itemize}
 \setlength\itemsep{-2pt}
  \item In “RECYCLE” mode, the cryo-cooler is recycling
  \item In “SETUP” mode, the instrument gets ready for observing (the filter wheel can be activated and calibration sources active)
  \item In “OBSERVATION” mode, the detection chain is active, the MXS are one and are running at a stable pulse rate and the cooling is active with regulation
\end{itemize}
 
\begin{table}[!h]
    \centering
    \begin{tabular}{|l|c|c|c|}
 \hline
    Unit & RECYCLE & SETUP & OBSERVATION \\
\hline
Digital Readout electronics (6 identical units)  & 718.2 & 718.2& 718.2\\
Instrument Control Unit & 20 & 20 & 20 \\
\cryoac\ & 28.8 & 28.8 & 28.8 \\
2K Core Controller & 34 & 28.8& 28.8\\
Filter wheel electronics & 15.8 & 27.9 & 17.0 \\
\hline
Total & 816.8 & 823.7 & 812.8 \\
 \hline
    \end{tabular}
    \caption{The electrical power budget for X-IFU in three instrument modes. The allocation for X-IFU is 1300 W, indicating that in all sizing cases, the margin is about 60\%. }
    \label{tab:electrical_power_budget}
\end{table}
\subsection{Science telemetry budget}
The six DRE and the \cryoac\ will produce science telemetry packets that are transmitted to the platform for storage before downlink to the ground. The DRE and CryoAC will implement output buffers and overflow control to ensure they do not exceed the permitted on-board data-rate. Each event requires 120 bits to be fully described. For a moderately bright X-ray source of 10 mCrab, this leads to a data rate of 75 kbits/s and a volume of 7.5 Gbits for a 100 ks observation. The X-IFU further produces a steady rate of 18 kbits/s, as the sum of the background bit rate, calibration bit rate, \cryoac\ bit rate (1.8 Gbits for 100 ks). These volumes are to be compared with the 26 Gbits of mass memory allocated daily to X-IFU. Larger volumes of data would still be allowed for the observations of brighter sources, provided adapting the WFI observations to produce lower data rate around this period. 


\section{Performance assessment}
Next we present the latest assessment of the main X-IFU performance parameters, starting with the instrument efficiency.
\subsection{Instrument efficiency}
\label{subsec_ie}
The instrument efficiency (IE) refers to the probability of a photon arriving at the top of the aperture cylinder in the X-IFU field of view to be detected by the instrument and incorporated in the final science data. This includes the absorber quantum efficiency and the array geometrical filling factor (see \S \ref{sec:tes_array}), the transmission of the thermal filters (see \S \ref{sec:filters}), the loss of transmission due to contamination, the loss due to dead time process, as well as the detector yield. This excludes however dead time due to MXS-related veto processes, which rather corresponds to calibration time, as well as the rejection of events at high count rates, which is captured in the throughput requirements. 


Molecular contamination onto the thermal filters will lower their transmission, notably at the lowest energies. We chose to allocate 20\% (End-of-Life) of transmission loss at 0.35 keV (corresponding to a transmission loss of 4\% at 1 keV). 
The most likely candidate species for contamination are carbon (hydro carbonate) and oxygen (water), which coincidentally represent the worst case species at 350 eV and 1 keV. Particulate contamination is deemed negligible.

\begin{table}[!h]
    \centering
    \begin{tabular}{|l|c|c|c|c|}
\hline  Energy &0.35 keV &1 keV &7 keV &10 keV \\
\hline 
Absorbers & 0.96 & 0.96 & 0.87 & 0.58 \\
Filters & 0.21& 0.76 & 0.89 & 0.89 \\
Contamination & 0.80 & 0.96 & 1.0 & 1.0 \\
Yield & 0.92 & 0.92 & 0.92 & 0.92 \\
Deadtime & 0.98 & 0.98 & 0.98 & 0.98 \\
\hline 
Total & 0.143 & 0.629 & 0.695 & 0.463 \\
Requirement & 0.13 & 0.57 & 0.63 & 0.42 \\
Margin & 10.2 \% & 9.4  \%& 9.4  \% & 9.4 \% \\
\hline
\end{tabular}
    \caption{Instrument efficiency allocation.}
    \label{tab:ie_allocation}
\end{table}
The detector yield is expected to be dominated by the fraction of ``dead pixels'' in the sensor array (due to various fabrication defects). Accounting for all the possible losses in the different cold connections, as well as the MUX SQUID chip fabrication yield and possible loss via ESD events, we expect a total yield of $\sim 92$\%, to be consolidated after testing all fabrication processes on large numbers of samples.

 As for dead time, we currently identify two  sources, \cryoac\ vetoes and thermal background, for a total of 2\% equally split between these two contributors. 

\begin{figure}[!ht]
    \centering
    \includegraphics[width=10cm]{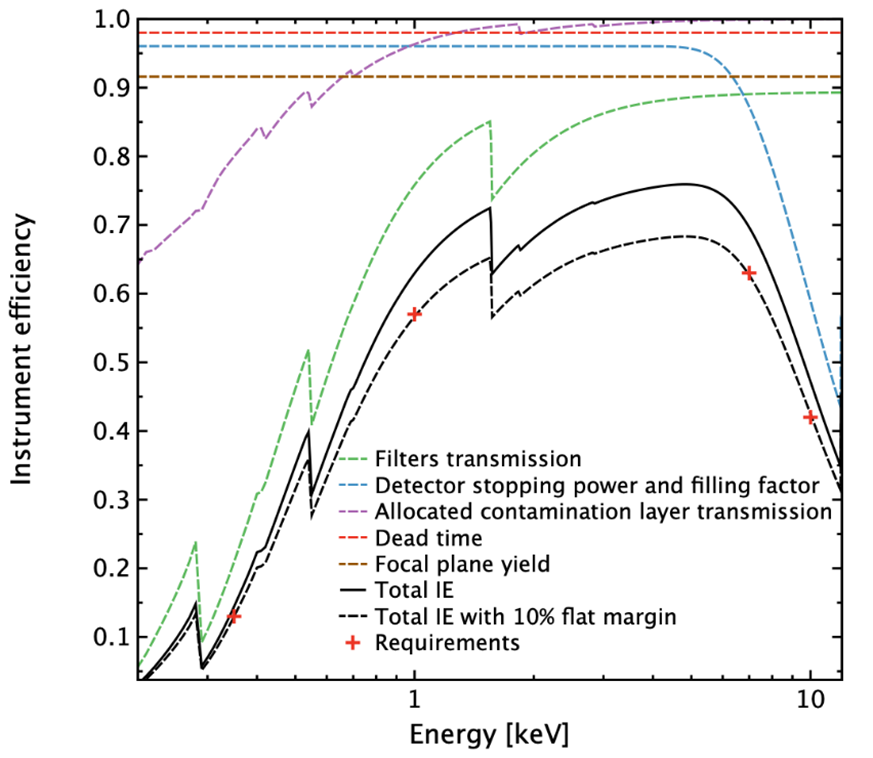}
    \caption{The variation of the instrument efficiency with energy, according to the breakdown of Table \ref{tab:ie_allocation}.}
    \label{fig:instrument_efficiency}
\end{figure}

The breakdown of the instrument efficiency following the assumptions listed above is shown in table \ref{tab:ie_allocation}. As can be seen, the present allocations leave a margin of $\sim 10$ \% at all energies (see Figure \ref{fig:instrument_efficiency}). This reserve may notably be used to compensate for lower than expected TES array yield or to allow for a safer filter mesh design ($\sim 5$\% higher blocking factor).

\subsection{Energy Resolution}
The X-IFU energy resolution not only depends on the performance of its cold detection chain, but also on the effect of its environment. The current X-IFU resolution budget is dominated by the detector array intrinsic performance (2.1 eV of the total allocated 2.5 eV FWHM). The remainder (1.4 eV - contributions to the ERB add in quadrature) consists of $\sim 60$ contributions, each of which represents in itself a small increase of the total the instrument resolution (typically 0.1 - 0.2 eV) and must be carefully sized following our growing understanding of the different contributors and their criticality. Today, the energy resolution budget can essentially be separated into three types of contributors: 1) noise terms generated by or picked-up from the different elements of the readout chain, including the detectors, 2) low frequency gain drifts of the environment of the detectors and electronics, 3) higher order effects that cannot be associated to either category \cite{denHartog_2014SPIE.9144E..5QD,Peille_2018SPIE10699E..4KP}. All contributors (in eV) are added in quadrature to the total budget. The main contributors to the energy resolution budget are listed in Table \ref{tab:erb}.


\begin{table}[!h]
    \centering
    \begin{tabular}{|p{4cm}|p{3cm}|p{8.5cm}|}
\hline
Contribution & Allocation & Comment \\
\hline
$\Delta$E at instrument level & 2.5 eV & up to 7 keV \\
\hline Detector array & 2.1 eV  & Including non-linearity and non-optimal set points. Allows 1.4 eV for all other terms \\
Instantaneous gain calibration accuracy & 0.3 eV & To be achieved over 4000 s, and taking into account MXS limitations\\ 
Active crosstalk & 0.20 eV & Taken separately in lieu of a detailed budget for individual units \\
Focal Plane Assembly & 0.58 eV & The SQ1 receives a 0.45 eV contribution\\
Digital Readout Electronics & 0.57 eV & Incl. some extra margin for DACs. In particular the SQ1 feedback gets 0.46 eV. Arrival phase variation receives 0.2 eV. \\
Detector Cooling System & 0.59 eV & Incl. harness and aperture cylinder. Conducted susceptibility receives 0.22 eV \\
Warm Front-End Electronics & 0.46 eV & LNA white and $1/f$ noise receives 0.36 eV \\ 
Unspecified margin & 0.71 eV & \\   
\hline \end{tabular}
\caption{Main contributors to the energy resolution budget, leading to a 2.5 eV resolution at system level.}
    \label{tab:erb}
\end{table}

\begin{figure}[!ht]
    \centering
    \includegraphics[width=17cm]{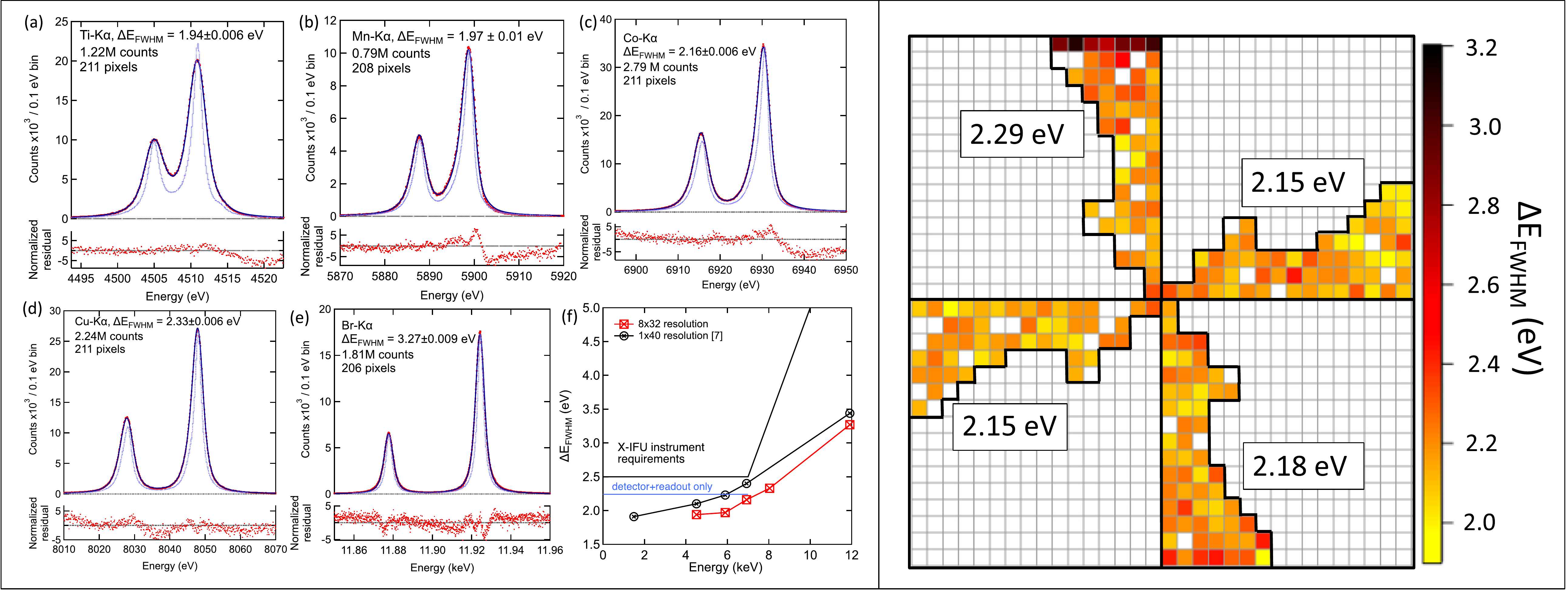}
    \caption{Left) From \cite{Smith_2021ITAS...3161918S} Co-added 8-column by 32-row spectra for (a) Ti-K$\alpha$, (b) Mn-K$\alpha$, (c) Co-K$\alpha$, (d) Cu-K$\alpha$ and (e) Br-K$\alpha$ (measured from K-Br). The red dots are the data points, the light blue lines are the natural lines shapes and the dark blue line is the ﬁt to data. The ﬁtted $\Delta$E (Full Width at Half Maximum, FWHM), number of counts and number of pixels included in each spectrum are shown on the ﬁgures (f) $\Delta$E FWHM  as a function of energy for the 1-column by 40-row measurements\cite{Durkin_2019ITAS...2904472D} and the new 8-column by 32-row results (red squares). The improved $\Delta$E FWHM is attributed to lower noise, improvements in the dynamic behavior of the readout chain and more optimized TES bias point. The dashed black line shows the X-IFU instrument level resolution requirements, and the dashed blue line shows the currently assumed requirements considering only the detector and readout subsystem and excluding margin. Right) From \cite{Smith_2021ITAS...3161918S} $\Delta$E FWHM array heat map measured at Co-K$\alpha$.}
    \label{fig:TDM_Spectral_resolution_smith_2021}
\end{figure}

As shown in Figure \ref{fig:TDM_Spectral_resolution_smith_2021}, TDM multiplexed, multi-channel readout has been already shown to meet the energy resolution requirement for X-IFU, using lab electronics\cite{Smith_2021ITAS...3161918S,Durkin_2019ITAS...2904472D}. 

\subsection{Energy scale knowledge}
\label{sec:escale}

The energy scale is the function that relates the output of the on-board Event Processor (in digital units) to actual calibrated keV units. The X-IFU  requirement on its knowledge is to achieve a 1$\sigma$ error of $<0.4$ eV in the range $0.2 - 7$ keV, over a 4 ks (TBC) timescale. This requirement derives from the need to measure accurately (down to 20 km/s accuracy) bulk motions of hot plasma in galaxy clusters, being deemed sufficient to measure the thermal broadening in typical clusters so to reveal any significant non-thermal pressure associated with gas motions in addition to bulk velocities\cite{Ettori_2013arXiv1306.2322E}.

Owing to the high sensitivity of the microcalorimeters, the energy scale function is sensitive to the different instrumental conditions. The real detector energy scale will thus drift over time and the use of a single ground-calibrated energy scale function would inevitably result in large systematic errors in the energy measurement. The present strategy is to accurately map during ground calibration the gain scale behaviour as a function of the main drift sources (temperature, magnetic field, optical load, etc...) and later interpolate the current gain scale using different real-time observables. For this, we will primarily use on-board modulated X-ray sources (see \S \ref{sec:cas}) providing X-ray photons at known characteristic energies, but also dedicated pixel readings and subsystem housekeeping data \cite{Porter_2016JLTP..184..498P,Cucchetti_2018SPIE10699E..4MC}. 

We identify four main contributors to the energy scale:
\begin{itemize}
 \setlength\itemsep{-2pt}
    \item the accuracy of the ground calibration of this function, limited by the statistics of the calibration data (0.1 eV 1$\sigma$),
    \item the instantaneous precision of the gain drift measurement that will be limited by the number of useful MXS counts available on the correction timescale (4 ks),
    \item the systematic non-linearities introduced by the imperfect settling of the readout signal at the end of a TDM row\cite{Durkin_2021ITAS...3165279D} (expected to be $<0.05$ eV with the X-IFU readout bandwidths),
    \item the finite accuracy of the gain drift correction algorithms and housekeeping data which introduce systematic correction errors. 
\end{itemize}

An extensive demonstration campaign is ongoing at GSFC using a prototype 8x32 TDM setup\cite{Smith_2021ITAS...3161918S} to demonstrate our ability to accurately correct gain drifts inside the expected parameter space of X-IFU operating conditions with different correction algorithms. First results indicate error levels relatively consistent with predictions obtained through simulations of the instrumental readout via XIFUSIM\cite{Kirsch_2022JLTP..tmp...59K,Cucchetti_2018SPIE10699E..4MC}, on which the instrument energy scale budget is based. Overall, our current best estimate for the energy scale performance amounts to 0.33 eV.

\subsection{Count rate capability}
In this section, we review the count rate capability of X-IFU, starting with the list of requirements to be met (see the last column, Table \ref{tab:count_rate_and_requirement}). One of the most emblematic count rate requirements for X-IFU is the one related to mCrab intensity sources observed out-of-focus requiring a throughput of 80\% for 2.5 eV spectral resolution, and is driven by the capability to observe bright gamma-ray burst afterglows to probe the Warm Hot Intergalactic Medium\cite{Brand_2016SPIE.9905E..5FB,Walsh_2020A&A...642A..24W}. Another one is certainly the one related to the capability of observing very bright X-ray sources (1 Crab intensity level) with X-IFU, thanks to the optics defocussing and the addition of thick filters along the light path. This one is driven by the objective to probe stellar mass black hole spins, winds and outflows in binary systems\cite{Motch2013arXiv1306.2334M}.
\begin{table}[!h]
    \centering
    \begin{tabular}{|p{4.15cm}|P{4.5cm}|P{2.75cm}|P{2.45cm}|}
\hline
Observation type & Astrophysical flux & Count rate at X-IFU TES array & Resolution \& (Throughput) \\
\hline
Defocused point source & 1 mCrab &62 cps & 2.5 eV ($>80$\%)\\
Defocused point source (G)  & 10 mCrab&620 cps & 2.5 eV ($>80$\%) \\
Brightest point source & 1 Crab&8.2 kcps (Be filt)& 10 eV ($>50$\%) \\
Focused point source & 0.25 mCrab & 15.5 cps &2.5 eV ($>80$\%)\\
Extended source &Perseus inner 1’x1’ &0.44 cps/pixel&2.5 eV ($>80$\%)\\
Extended source (G) &CasA average flux& 2.5 cps/pixel&2.5 eV ($>80$\%)\\
\hline
\end{tabular}
    \caption{This table lists the observation type (focused, defocused, point or extended source, G refers to the Goal), the astrophysical flux, the count rate at the X-IFU entrance, the X-IFU at the TES array, i.e. after passing through the filter wheel filter, and finally the requirement in terms of spectral resolution (in eV) and throughput (\%). The defocusing depth assumed is 35 mm, see Kammoun et al.\cite{Kammoun_2022arXiv220501126K}, for more details, including a method to analyze bright source defocused observations. For the requirements, the count rates are estimated assuming a Crab like spectrum for point source and a thermal APEC 4 keV absorbed APEC model for extended source. The reference telescope design used in this table is: ATHENA - Telescope Reference Design and Effective Area Estimates, ESA-ATHENA-ESTEC-PL-DD- 001, Issue 3.3, 21/12/2020. Note that the requirement of 10 eV resolution with a 50\% throughput applies to the 5 to 8 keV band, and is met assuming a 100 $\mu$m thick beryllium filter cutting off most of the photons below 2-3 keV.}
    \label{tab:count_rate_and_requirement}
\end{table}

We define the throughput as the fraction of valid events selected to pass a certain resolution criterion (currently 2.5 eV or 10 eV depending on the requirement) among all detected events. Most count rate requirements are thus expressed as a minimum throughput to be met at a given reference instrument count rate. This stems from the fact that the energy resolution of a photon detected by a micro-calorimeter depends on the time separation between individual X-ray pulses in the pixel timeline, as well as on the presence of crosstalk signals generated by simultaneous events on neighbour pixels: the 2.5 eV resolution will only be achieved for photons well isolated within the data stream timeline. 

\subsubsection{Components of the throughput budget}
The throughput performance depends on several factors\cite{Peille_2018JLTP..193..940P}. The main ones are listed below: 
\begin{itemize}
 \setlength\itemsep{-2pt}
    \item The instrument efficiency as described above, 
    \item The pixel pitch which drives the event spread of the sources onto different neighbouring pixels: smaller pixels would each receive lower count rates and better share the overall count rate load,
    \item The pixel speed. At high count rates, the current pulses used for the energy reconstruction of the impacting X-rays will get packed together in the pixels timelines, leading to two main degradations of the instrument performance. For each event, if the preceding pulse is too close, the energy reconstruction will be biased (by an amount depending on the time separation) and the event will need to be rejected from the science data. Such events are called secondaries. A nearby subsequent pulse will in turn shorten the amount of data available for the reconstruction and degrade the energy resolution. To characterize these effects, different grades can be defined as shown in Table \ref{tab:grades}.
    \item Another main limitation to the instrument count rate capability is crosstalk between pixels. Crosstalk can be defined as any science signal, originating from the absorption of energy on one pixel, contributing to an energy measurement on another pixel. It may arise via various mechanisms: 1) thermal leakage on the detector array, 
    2) next-in time crosstalk related to the imperfect settling of the TDM switch at the end of a row period, 
    and 3) inductive coupling due to the finite mutual inductance between elements of the readout chain.
    \item Event Processor (EP) performance. Due to limited on board computing power and memory, the X-IFU EP will have a maximal event processing rate. However, together with the processing of the MXS events, the EP is being designed for a maximum processing rate of $\sim$20 kcps, in order not to introduce further throughput losses.
    

\end{itemize}

\begin{table}[!h]
    \centering
    \begin{tabular}{|l|c|c|c|}
\hline
Grade & Time since  & Record length  & Resolution  \\
 & previous pulse & (time between the start&for primary events \\
 & & of the record and  & \\
 & &  the next pulse) & \\
 \hline
Very high resolution & $> 3676$ samples (20 ms) & $> 8192$ samples (44.6 ms) & 2.5 eV \\
High resolution &  $> 3676$ samples (20 ms) & $> 4096$ samples (22.3 ms) & $\sim 2.5$ eV \\
Intermediate resolution & $> 1838$ samples (10 ms) & $> 2048$ samples (11.1 ms) & $2.6$ eV \\
Medium resolution & $> 1838$ samples (10 ms) & $> 512$ samples (2.8 ms) & $3.0$ eV  \\
Limited resolution & $> 1838$ samples (10 ms)& $> 256$ samples (1.4 ms) & $7.0$ eV  \\
Low resolution & $> 1838$ samples (10 ms)& $> 8$ samples (44 $\mu$s, TBC) & $\sim 30.0$ eV \\
Secondaries  & $> 1838$ samples (10 ms)& - &  N/A \\
Invalid  & - & $< 8$ samples ( 44 $\mu$s, TBC) &  N/A\\
 \hline
    \end{tabular}
    \caption{Definition of the different event grades for X-IFU.}
    \label{tab:grades}
\end{table}

\begin{figure}[!t]
    \centering
    \includegraphics[width=15cm]{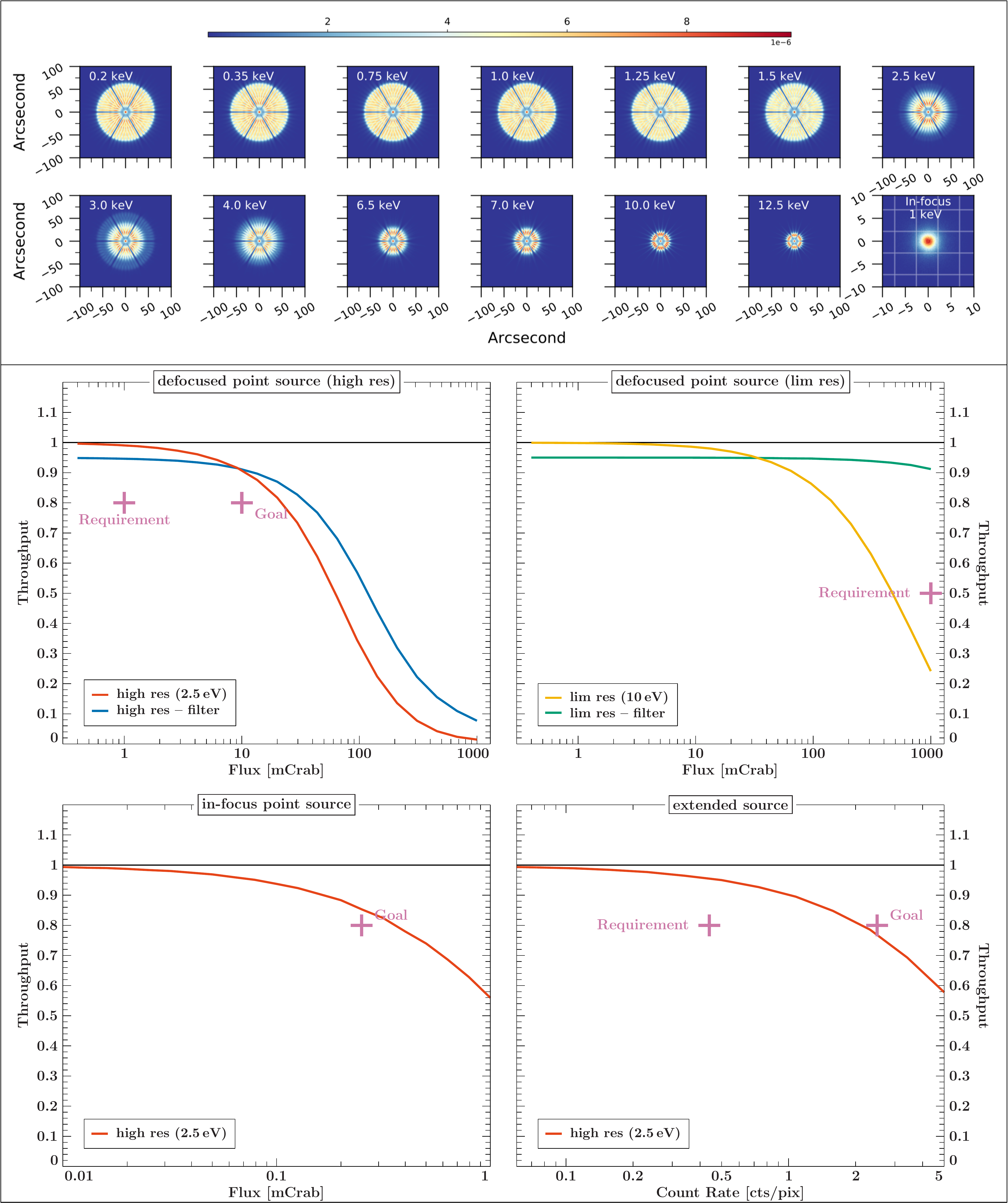}
    \caption{Top: The defocused PSF of the Athena optics  at 13 energies as seen by X-IFU\cite{Kammoun_2022arXiv220501126K}. The bottom rightmost panel shows the in-focus PSF at 1 keV (Half Energy Width = 5"), for comparison. The grid in this panel shows the X-IFU pixels. We note that the in-focus PSF is shown on a 20" × 20" image while the defocused PSF is shown on a 200" $\times$ 200" image. The full field of view of the X-IFU is of the order of 300" (equivalent diameter). Bottom: Summary of the throughput performance for all the observing cases specified in the X-IFU URD. Requirements and goals are shown with crosses. Top left: High resolution observation of a defocused point source. Top right: Limited resolution (10 eV) observation of a defocused point source (for the filter configuration, the plot includes the 5 \% loss due to the limited transmission of the beryllium filter in the 5 to 8 keV band of interest). Bottom left: High resolution observation of a focused point source. Bottom right: High resolution observation of an extended source.}
    \label{fig:throughput}
\end{figure}

Figure \ref{fig:throughput} shows the expected instrument throughput performance as a function of count rate/flux for the different observation types specified in the X-IFU count rate requirements, and relying on the defocusing capability of the Athena telescope (see Figure \ref{fig:throughput} to visualize how the point spread function of the optics varies with energy at an out-of-focus distance of 35 mm, \cite{Kammoun_2022arXiv220501126K}). These results were obtained using detailed End-to-End simulations performed in the SIXTE environment\cite{Dauser_2019A&A...630A..66D}, taking into account all the above-mentioned processes\cite{Peille_2018JLTP..193..940P}. As can be seen, a significant margin exists with respect to all performance requirements, and almost all goals are met. The only exception is the extended source goal by a few percentage points. This actually shows that the X-IFU instrument is not being oversized. This study notably serves as an input to key subsystem performance requirements, in particular those related to pixel speed constraints and crosstalk level.

\subsubsection{Undetected pileup}
Pileup originates at high count rates when events are too close to one another in a pixel timeline, such that the triggering stage of the event processor will only detect one event. Phase A studies showed via simulations of the X-IFU trigger algorithms a fraction of undetected pileup events of less than 0.1 \% for all the X-IFU high count rate cases, i.e. a factor of 10 margin with respect to the requirement \cite{Cobo_2018SPIE10699E..4SC}. This is achieved thanks to the sharp rise of the X-ray pulses, which can be be easily identified in the low-noise pixel timeline. 

\subsection{Non X-ray background}


 The non-X-ray background includes all events generated by charged particles depositing energy in a pixel and in the instrument energy range that make it to the scientific data\cite{Lotti_2021ApJ...909..111L}. Its level is defined after all veto mechanisms have been applied (currently includes coincidence with a \cryoac\ event and coincidence with another event in the array to reject secondary showers - see \S \ref{sec:cryoac}). 



The X-IFU background rejection efficiency depends on the geometrical configuration of the \cryoac\ and main detector array - related to the classical concept of the anti-coincidence solid angle coverage -, and on the capability of the main array to discriminate background events on its own, relying on the energy deposited and on the pixel pattern turned on by the impacting particles (“detector rejection efficiency”). Two different screening strategies are thus applied to the flux incident on the X-IFU:
\begin{enumerate}
    \item Time coincidence with the \cryoac: photons are not expected to produce a coincidence signal in the \cryoac, so we can assume that every event that is detected simultaneously in both the \cryoac\ and the main detector can be rejected as particle-induced one.
    \item Pattern recognition: due to the detector features (no charge cloud diffusion among different pixels as in CCD-like detectors, pixels physically separated) source photons will not produce complicated pixel patterns to reconstruct. Instead, we can assume that all the events that turn on more than one pixel can be rejected as induced by particles with skew trajectories intersecting more than one pixel, or by simultaneous impacts by multiple particles.
\end{enumerate}

Extensive Monte Carlo simulations have been performed using the Geant4 software to assess the X-IFU background level for different geometrical  configurations\cite{Lotti_2021ApJ...909..111L}. Special care was taken to incorporate the most representative mass model of the FPA. These simulations have shown the need for a low Z coating (e.g. kapton) inside the FPA Nb shield to limit the secondary electron production in close proximity of the detector.  
Our current best estimate of the instrumental background, accounting for a previously used reference configuration for the FPA indicates a non-compliance of the order of 26\% compared to the requirement ($5 \times 10^{-3}$ counts  cm$^{-2}$ s$^{-1}$ keV$^{-1}$, 2-10 keV band, excluding fluorescence photons), taking into account a 20\% modeling margin based on a comparison exercise performed between Geant4 simulations and the measured in orbit Hitomi/SXS instrumental background\footnote{Ozaki \& Fioretti 2018: \url{https://indico.esa.int/event/249/contributions/4195/attachments/3262/4235/Hitomi-related_Geant4_activities.pdf}}. However, with an updated FPA mass model (e.g. with the location of the 50 mK filter at the bottom of the Niobium shield and larger clearance from the detector surface) it appears that the non-compliance may reach around 40-50\% including the modeling margin. We recall these estimates are based on worst case assumptions for the L1 environment (solar minimum), such that much lower background periods can be expected along the mission lifetime.

\begin{figure}
    \centering
    \includegraphics[width=17cm]{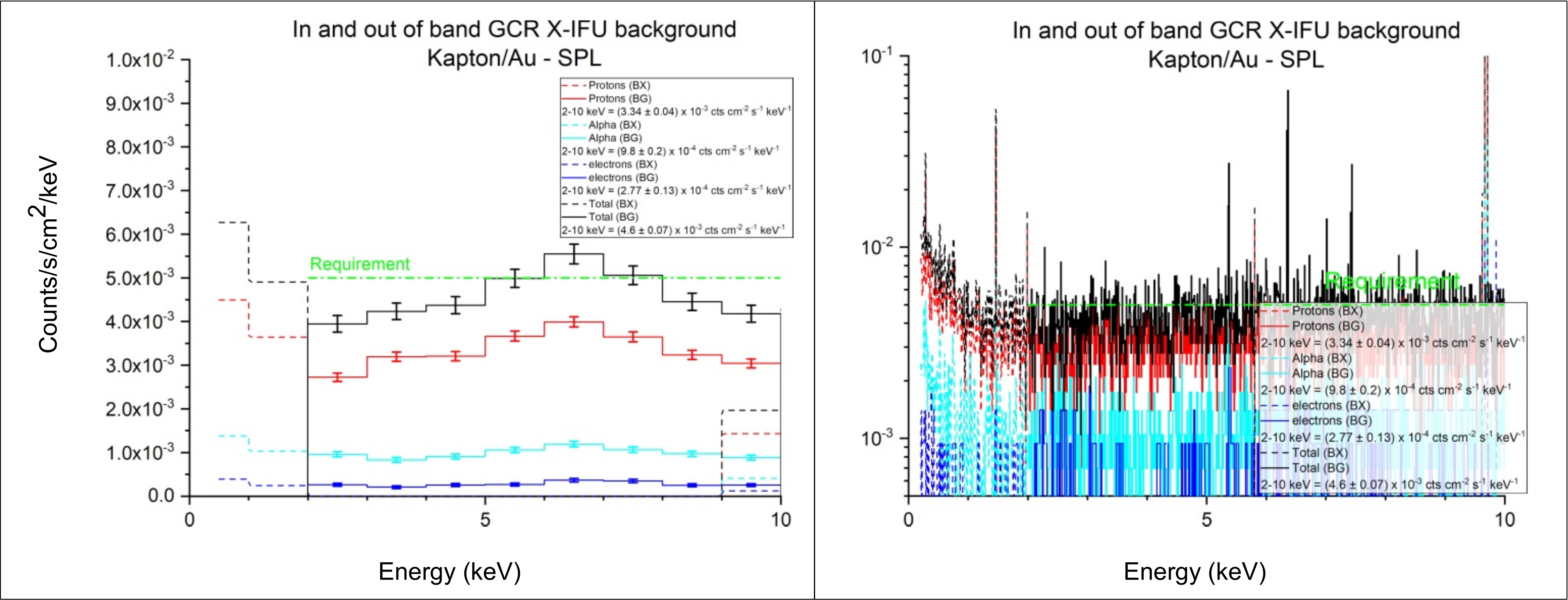}
    \caption{Total background level (GCR protons, $\alpha$ particles and electrons) of $4.6 \pm 0.07 \times 10^{-3}$ counts cm$^{-2}$ s$^{-1}$ keV$^{-1}$, (no margin included), at a resolution of 1 keV on the left and 10 eV on the right. These 100 ks simulations were performed on the Kapton/Au passive shielding configuration, with the 0.25 mm clearance from the detector surface. Note that the gold fluorescence peak at 9.7 keV is clearly visible.}
    \label{fig:background}
\end{figure}
\subsection{Targets of Opportunity (ToO)}
Concerning ToO, X-IFU has to fulfill two scientific requirements stated as : 
\begin{itemize}
 \setlength\itemsep{-2pt}
 \item X-IFU shall be able to deliver 50 high-resolution X-ray spectra of GRB afterglows with at least 1 million counts over its nominal mission lifetime.
    \item X-IFU shall be able to perform observations of a ToO in four hours wrt. ToO alert with at least 50 ks (TBC) available exposure for 40\% (TBC) of the events (assuming a 50\% Field-of-Regard, FoR)
\end{itemize}
The X-IFU, with its cooling cycle, plays a key role in the ToO response capability, as a successful observation is only possible if sufficient cold time is available at the appropriate time. Of course, in the end, the corresponding performance depends on the full Athena system. 
The main instrument contributors are: 1) The cold time  (a significant fraction of “successful” ToO are observed without recycling the instrument), 2) The duty cycle (ratio between the cold time and the sum of the cold time and regeneration time), and 3) The duration of a detector set-up, including possible calibrations required to get the right performance at the beginning of the observation. 

To optimize the instrument fast response, different recycling schemes have been investigated via extensive Monte Carlo simulations, accounting for a full modeling of the Athena system and its mock observing plan\cite{Jaubert_2019}. The best strategy (so called “partial heat-up”) uses the capability of the sub-Kelvin cooler to anticipate the early phase of the recycling (heat-up) and wait for a variable time (standby), before finishing the recycling and cooling down. The reference tuning of the cooler is a regeneration time of 8h50, a cold time = 28h30, corresponding to a duty cycle = 76.3 \% (note that a conservative 30 minutes of cold time is reserved prior to any observation to properly set up the instrument).

A sensitivity analysis of the ToO requirement with respect to the hybrid cooler performance parameters (cold and regeneration times) was also performed. This showed that the requirement expressed as 4 h delay / 50 ksec of observing time for 80\% of accessible ToO (FOR = 50\%) is never met for all cases of tuning. Instead a 8 h delay / 50 ksec is achievable for the “reference tuning” of the cooler. For the requirement expressed in terms of number of photons collected and number of successful ToO over the mission lifetime, this objective is always met with significant margins for all tuning of the X-IFU cooler. For the reference tuning, X-IFU could record up to $\sim 25$ one million count GRB afterglow spectra per year, indicating a factor of 2 margin in performance. The Athena effective observing efficiency was further shown not to depend at first order on the cold time of X-IFU. On the other hand, to optimize the Athena scheduling performance, the X-IFU duty cycle should be larger than the fraction of time allocated to X-IFU.

 \section{Technology demonstration and schedule}
 A demonstration plan of critical technologies and design has been elaborated in view of the mission adoption and later the instrument Preliminary Design Review. This plan includes demonstration activities at different levels, from key technological technologies to the coupling of several subsystem demonstrators. 
 
 At subsystem level, the items considered in the plan are :
 \begin{itemize}
  \setlength\itemsep{-2pt}
  \item the FPA and associated technologies, e.g. the TES array, the Kevlar suspensions, the SQUIDs, internal harnesses\ldots
  \item the hybrid cooler, e.g. the ADR superconducting coil\ldots
  \item the \cryoac, e.g. the \cryoac\ TES chip and CFEE\ldots
  \item the thermal filters, e.g. their resistance to vibration loads\ldots
  \item the aperture cylinder, e.g. EMC tightness, contamination protection\ldots
  \item the readout electronics, e.g. for the DRE: the RAS and DEMUX modules, the DC/DC converter, and for the WFEE: the ASIC holding the low-noise amplifier\ldots
  \item the cold harnesses, e.g. technology trade-off between looms and shielded twisted pairs\ldots
  \item the optical filters of the filter wheel
 \end{itemize}
Not all sub-systems are involved in the plan, as some have already reached or exceeded the required level of technology readiness. Those activities are mostly funded by the national funding agencies, with support from ESA in some cases.
 
 At system level, there are two main activities foreseen: 
 \begin{itemize}
  \setlength\itemsep{-2pt}
  \item The first one involves the first demonstration of a 2K Core 1) to validate the integration of the FPA DM with the subKelvin cooler (EM model from Safari) and 2K core structure, 2) to assess the sensitivity of the 2K core performances against external perturbations (microvibrations, thermal stability, magnetic field), 3) to characterize the recycling phase and stabilization time needed after recycling phase, and finally 4) to determine the best way to thermally regulate the 50mK stage (=T0 stage) of FPA DM. This activity is funded in part by ESA in the context of the Core Technology Program "Detector Cooling System", and is supported by JAXA through the key procurement of a 2K Joule Thomson cooler\cite{Prouve_2020Cryo..11203144P,Shinozaki_2019MS&E..502a2069S}. The activity will rely on a demonstration model for the FPA, holding an array of 80 pixels readout by a frequency domain multiplexing lab electronics (note that TDM readout was selected as the X-IFU baseline after the design activities of the FPA DM had started).
 \item The second one involves the demonstration of a multiplexed readout chain featuring design solutions for X-IFU (e.g. balanced differential readout, impedance matching, filtering at 2K, etc...), which were put in place to mitigate two risks: EMI/EMC compatibility and the impact of parasitic couplings on the dynamic behaviour of the TDM signals. The intention is to validate early on these design choices, as well as the associated interfaces between the different elements of the chain, in particular well in advance of the EM test campaign. 
 This activity constitutes the so-called early verification phase. It will rely on the development and coupled testing of demonstration models of the X-IFU electronics (WFEE + DRE), as well as breadboards to study the electrical interface between the FPA, WFEE and cold harness. The demonstration electronics will eventually be coupled to a dedicated cryogenic test platform, housed in a commercial two-stage ADR cryostat, hosting a focal plane array placed at 50 mK with a kilopixel TES array and associated cold readout electronics\cite{Betancourt_2021arXiv210703412B} (see Figure \ref{fig:gse_cryostat} left). 
 As discussed above, the choice to implement TDM readout for X-IFU was based mainly on the maturity level of this solution. Demonstration in laboratory set-up have shown performances compliant, if not exceeding X-IFU requirements\cite{Smith_2021ITAS...3161918S,Durkin_2019ITAS...2904472D}. 
 \end{itemize}
 \begin{figure}[!h]
     \centering
     \includegraphics[width=17cm]{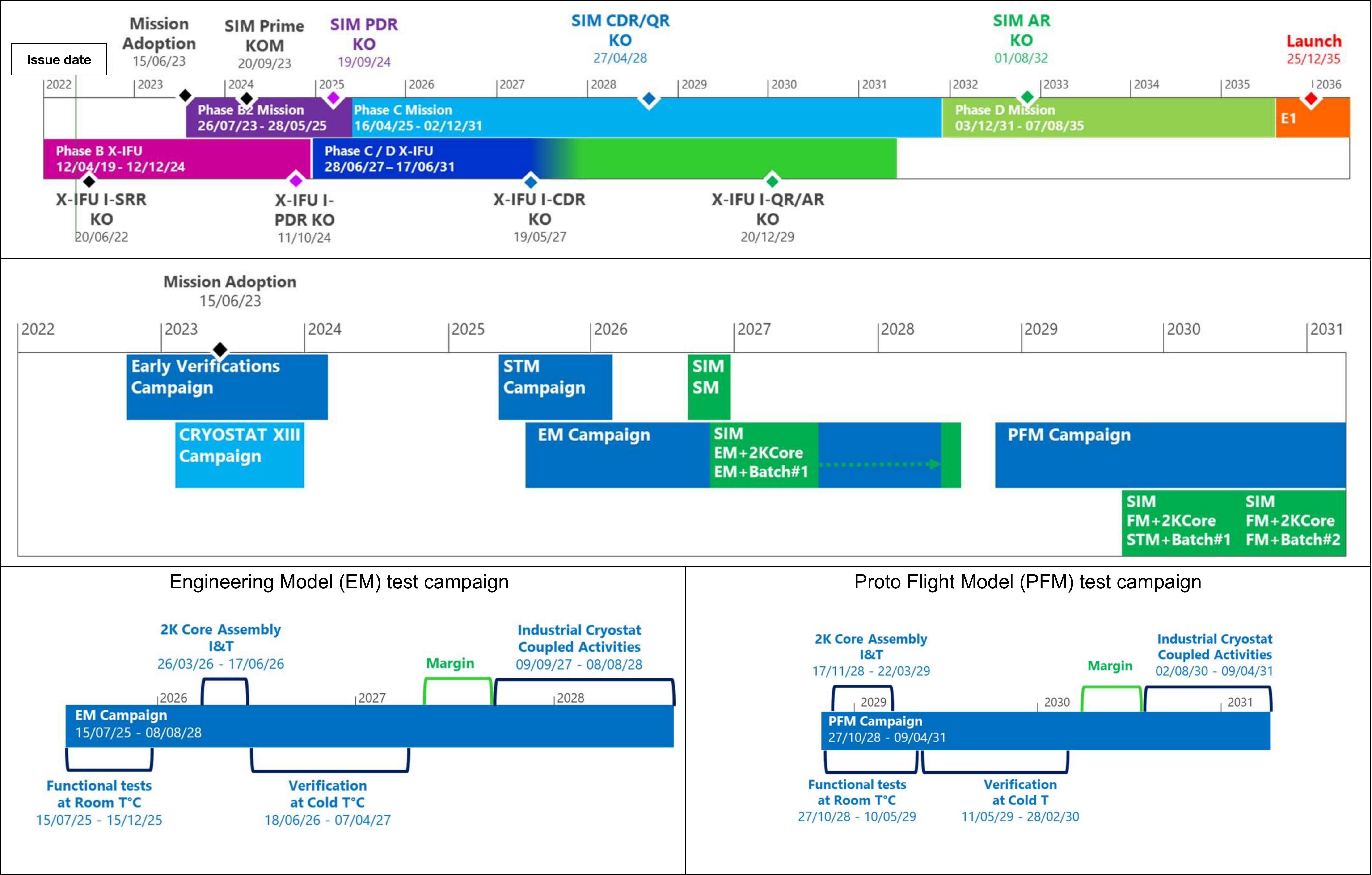}
     \caption{Top panel) The main instrument/mission development phases. Mid panel) The main instrument test campaigns. Cryostat XIII refers to the on-going DCS development. Bottom panels) On the left the engineering model test campaign and on the right the proto-flight model test campaign.}
     \label{fig:schedule}
 \end{figure}
 
 At the time of the SRR preparation, the assumption for the launch date of Athena was the end of 2035  and the adoption of the mission in June 2023. The development schedule of X-IFU, including the early verification phase, as iterated with ESA is presented in Figure \ref{fig:schedule}. Although this schedule is now obsolete, it is worth noting that the X-IFU engineering model delivery to industry was planned for September 2027, while the delivery of the flight model was planned by August 2030 (meaning about 3 years were needed to move from one model to the next). The models would have then been integrated in the cryostat at the Prime premises and coupled tests performed, including calibration activities. Note that these coupled activities at the Prime facilities are costly and will be part of the items to discuss as part of the upcoming design to cost exercise. The relatively long leap time separating the X-IFU model delivery dates and the launch date will also have to be looked at for optimizations.

\section{Calibration}
X-IFU being a sensitive high resolution spectrometer, special care is required for its calibration both on the ground and in-flight\cite{Cucchetti_2018SPIE10699E..4OC,Barret_2019A&A...628A...5B}. The X-IFU calibration strategy combines measurements and analysis: 
\begin{itemize}
  \setlength\itemsep{-2pt}
    \item At component level (detector array, filters, readout electronics, ...),
    \item At sub-system level (Focal Plane Assembly, detection chain, …),
    \item At instrument level (integrated in the GSE cryostat, on the Science Instrument Module, during Thermal Vacuum/Thermal Balance of the SIM and/or the spacecraft which should be the last opportunity to have a cold instrument before launch), 
    \item In-flight on internal sources or specific X-IFU configuration (ie on MXS, with filter wheel in closed position, specific acquisition mode during the calibration performance verification phase,...), 
    \item In-flight on sky sources, 
    \item X-ray emission fundamental physics modeling and ground measurements, at high energy resolution. These include simulations or revised and updated standards needed for such a high-resolution spectroscopy mission. 
\end{itemize}

\begin{figure}[!h]
    \centering
    \includegraphics[width=17cm]{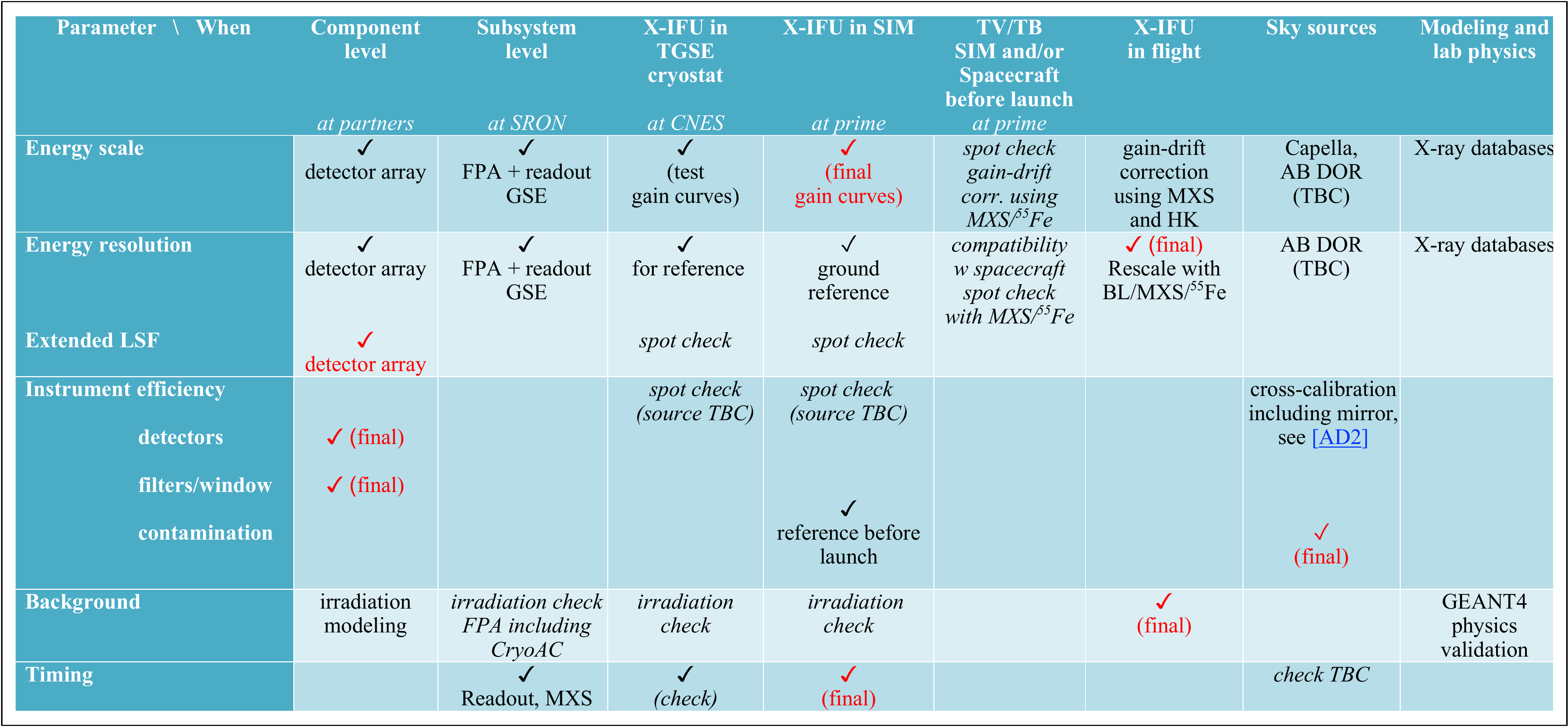}
    \caption{Calibration strategy outline.A check symbol \checkmark indicates when calibration activities are performed. Text entries indicate the component or sub-systems that are considered or other specific characterization or action. The red color highlights the last opportunity to calibrate the parameter. Italics refers to AIT/AIV activities.}
    \label{fig:calib_strategy}
\end{figure}

Figure \ref{fig:calib_strategy} gives the strategy that will be followed for the calibration of the various instrument parameters. As can be seen, most of the final ground calibration references are expected to be acquired at SIM level, i.e. in the flight cryostat of the X-IFU instrument in order to have a representative environment including all the major drivers of the performance. This calibration campaign is expected to last $\sim$ 4 months, a duration that is mainly driven by the need to acquire accurate reference energy scales under different operating conditions (see \S \ref{sec:escale}). 

\begin{figure}[!h]
    \centering
    \includegraphics[width=17cm]{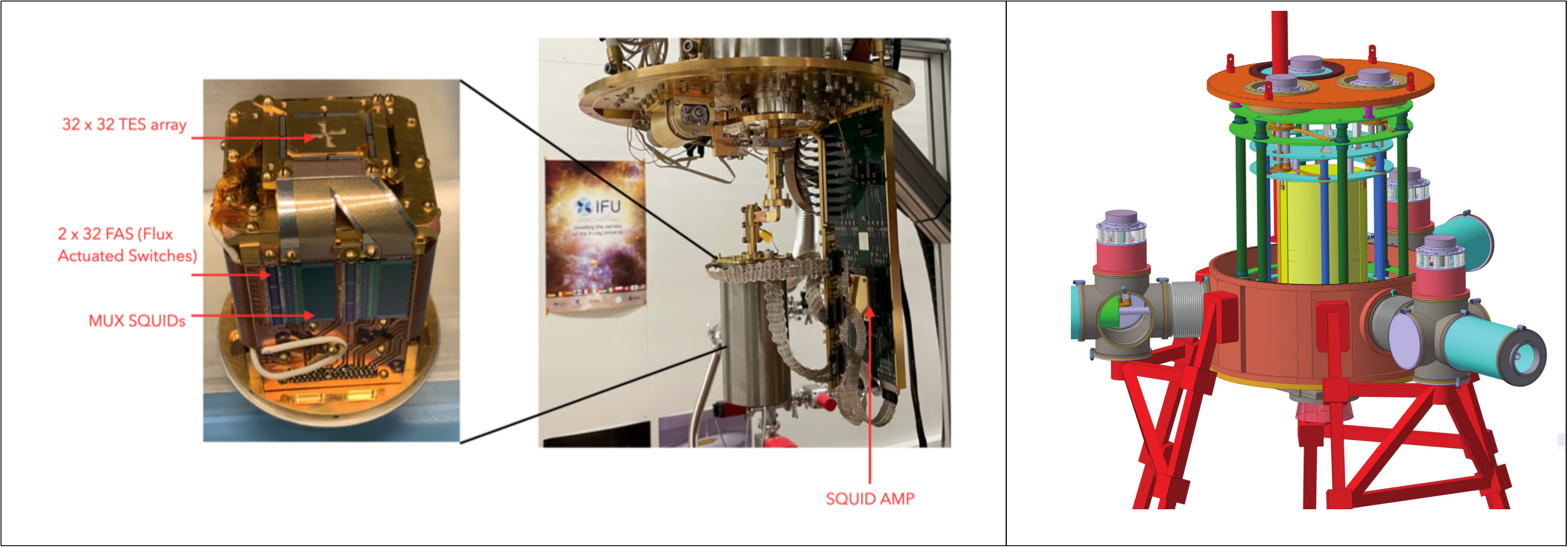}
    \caption{Left) The snout is composed of a 1024 TES array with its associated cold readout electronics. The snout is placed in the niobium shield and connected to the 50 mK stage. Superconducting looms insure the signals connection to the 500 mK terminator card and then to the 3 K cold electronics\cite{Betancourt_2021arXiv210703412B} (see Castellani et al., 2022, SPIE submitted). Right) Conceptual scheme of the cryostat for testing and calibrating the Focal Plane Assembly of Athena/X-IFU. Credit: INTA – C2CC team (and AthenaNuggets62). }
    \label{fig:gse_cryostat}
\end{figure}

One critical step of our verification and calibration is also the thermal GSE (also C2CC, for 2K Core Calibration, see Figure \ref{fig:gse_cryostat}, right). The C2CC is an assembly of structural, mechanical, electrical, and software elements designed to verify the functions and performances of the 2K core, under its operational conditions. This means that this cryostat should provide the adequate environment at the 2K core interfaces in terms of micro-vibrations, EMC (especially magnetic field), thermal (temperature level and thermal stability). An aperture cylinder including thermal filters GSE will be also integrated into the C2CC. A functional cold harness representative of the flight ones in terms of material, length, and thermal interfaces will be connected to the 2K core. This stage constitutes the last level of integration prior to the instrument delivery, allowing to verify the instrument performances, including their sensitivity to environment conditions and perform pre-calibrations in advance of the more extensive SIM level campaign. 

For the instrument calibration, a diversity of calibration sources will be used, as described below:
\begin{itemize}
  \setlength\itemsep{-2pt}
    \item First, we plan to use rotating target sources (RTSs) similar to the ones used for Hitomi/SXS\cite{Eckart_10.1117/1.JATIS.4.2.021406}. RTSs provide a rotating scan of the X-ray lines at a given frequency, which enables them to observe all line complexes in a single cold time and provides a way to track long term drifts in the ground calibration measurements.
    \item These rotating sources will be complemented by the use of a dedicated Electron Beam Ion Trap (EBIT,\cite{Levine_1988PhST...22..157L}). Many of the known L/K fluorescent lines at $<1$ keVs are indeed broad, and their detailed shape and centroid are either not accurately known, or can depend on chemical composition of the target and thus do not constitute reliable references. EBITs can produce narrower (typically tenth of eV FWHM) well-known X-ray lines at low energy, providing a more accurate description of the energy scale for a lower number of counts (approx. 5000 counts instead of 10000 for $K_{\alpha}$ lines). The use of this source has clear implications on the X-IFU interface due to its large volume, which will need to fit below the SIM. 
    \item Channel-Cut Crystal Monochromators (CCCMs)\cite{Leutenegger_2020RScI...91h3110L} will also be used for more efficient energy resolution measurements, but also to verify the calibration of the instrument extended line spread function performed at TES array level.
    \item A flight-like MXS and $^{55}$Fe sources that are used to reproduce, without using the flight hardware, the fiducial lines expected in-flight throughout the calibration.
\end{itemize}

Some calibration activities will be performed at system level. The extended line spread function will be calibrated at TES array level with CCCMs as it only depends on the physics of photon absorption in the detector. On the other hand, the X-IFU will not be exposed to a synchrotron beam to calibrate its quantum efficiency, given the complexity of such test. Instead we will rely on a set of measurements at sub-system level (thermal filters and TES array). 

The possibility of irradiating a representative FPA by protons was also investigated as a way to validate the models used for the computation of the expected NXB. This was not followed up in favor of more focused verification activities of the key processes dominating the X-IFU background (electron back scattering on XIFU-like absorber samples and secondary electron production in the kapton-coated Nb shield). For the calibration of the NXB, we plan to perform long dark exposures with the filter wheel in the closed position, further relying on a high-performance radiation monitor tracking variations of the charged particle environment. 

\section{Overview of the X-IFU Instrument Science Center (\xisc)}
\begin{figure}[!h]
    \centering
    \includegraphics[width=17cm]{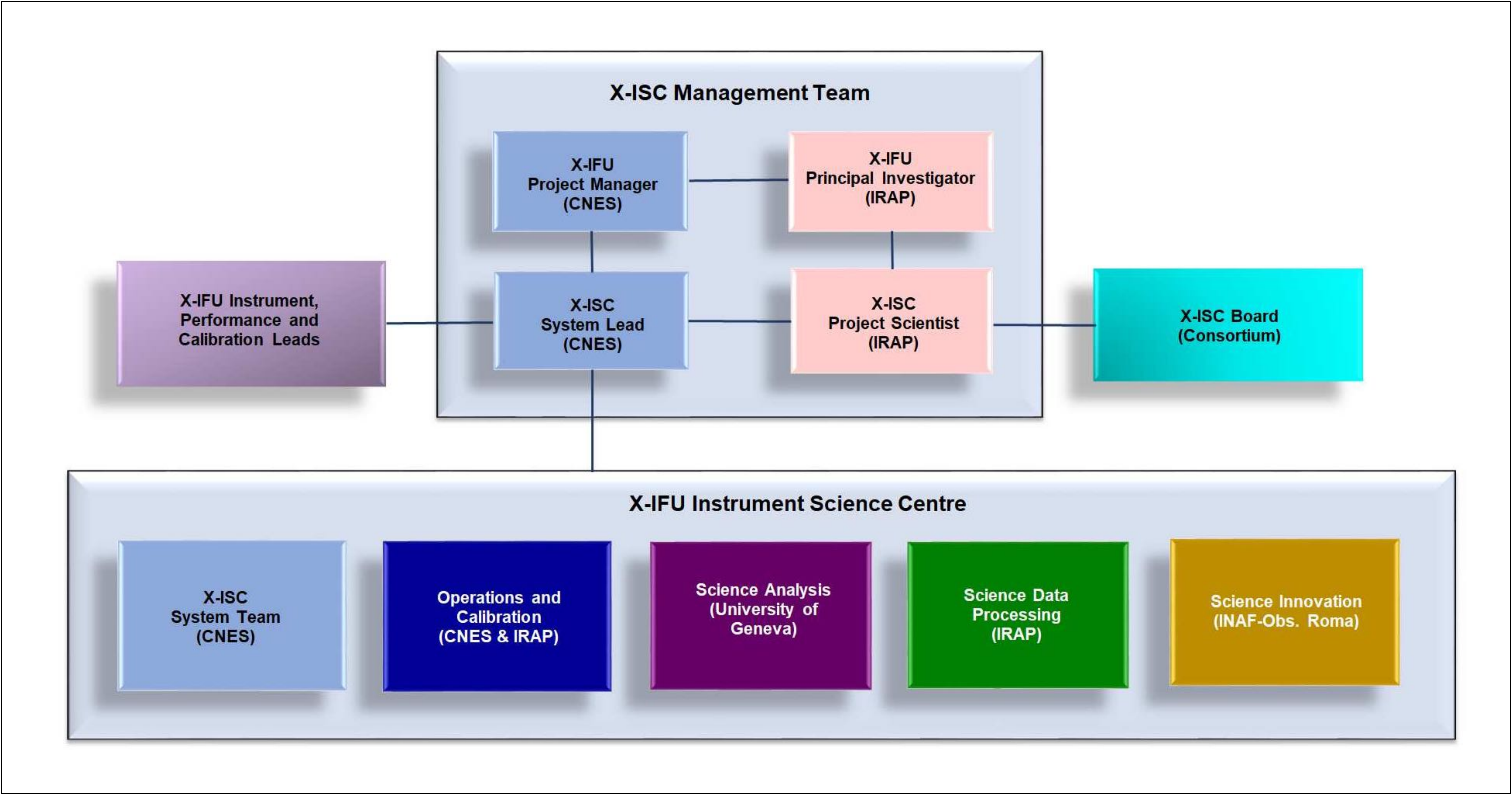}
    \caption{The overall organisation structure for the \xisc. Only the Prime contributors are shown. Other contributions from Consortium partners will make up the \xisc.}
    \label{fig:xisc}
\end{figure}

The X-IFU Instrument Science Center (\xisc) is an integral part of the instrument delivery. It will be developed within the X-IFU Consortium, with the activities ramping up 5 to 7 years before launch. Prime contributors to the \xisc\ are France, Switzerland, and Italy (see Figure \ref{fig:xisc}).  The \xisc\ includes responsibilities for X-IFU instrument operations and calibration. In conjunction, the \xisc\ will be responsible for instrument performance evaluation, instrument calibration and trend monitoring during the mission operations, contiguous with their role in instrument testing, characterisation and calibration pre-launch. The \xisc\ will generally be responsible for the provision of all detailed instrument knowledge and expertise to successfully operate the instrument. It will operate together with the ESA Mission Operation Center (MOC) and Science Operation Center (SOC), and form the X-IFU component of the Athena Science Ground Segment (SGS).
More specifically, in the current share of work with ESA as described in the science implementation plan, the main responsibilities of the \xisc\ are:
\begin{itemize}
  \setlength\itemsep{-2pt}
   \item Instrument testing, validation, characterisation and calibration, including definition of instrument tests to be performed in the in-flight Checkout, Commissioning and Performance Verification Phases.
   \item In-flight instrument performance evaluation, instrument calibration and trend monitoring and following-up instrument operational anomalies.

\item  Production of instrument and software user manuals, and inputs for the observers’ manuals and for all instrument specific user support documentation. In particular, the X-ISC will review and formally endorse instrument manuals issued in support of the calls for observing proposals.


\item  Provision of a scientific simulator and instrument observation-time estimators: tools to simulate planned observations and to calculate s/n vs. observation time and vice-versa, for use in proposal preparation as well as preparation of calibrations and any other observations/measurements. They are critical for an efficient planning and will need to be incorporated early within the SOC framework.


\item  While the SOC will be responsible for Athena Common Software System (ACSS) coordination, \xisc\ will allocate resources for a joint ACSS coordination board and join the SOC in the testing and validation of new releases of the ACSS; \xisc\ will provide S/W contributions and carry out ACSS related activities. These comprise (but are not limited to) the development of the data models, common libraries, pipeline scripting language, graphical user interfaces, as well as S/W integration activities and the development of S/W integration servers.

\item  Development and provision of the standard scientific data processing (DP) S/W and calibration files that are used both for the pipeline and interactive (IA) processing, including documentation. The IA scientific DP S/W, including necessary calibration files, will be distributed via the SOC.

\item  Development, testing, and operation of the scientific data processing pipeline to process Level-0.5 data into Level-1 and 2 products. Pipeline products will be delivered to SOC for storage and distribution.

\item  Production of value-added scientific data products, including catalogues, surveys, and other contributions to Level-3 products.

\item As part of the \xisc\ daily and long-term instrument performance monitoring, the \xisc\ will be responsible for the verification, quality control, and validation of all data products, including the quick-look analysis (QLA) of products suited for the health monitoring of the instruments. 

\item  Provision of up-to-date instrument calibration parameters (calibration files) 



\item  Assisting the SOC in supporting the user community in those cases where in-depth instrument expertise beyond SOC capabilities is required.

\end{itemize}

\section{X-IFU Consortium organisation}
The X-ray Integral Field Unit (X-IFU) and its associated X-IFU Instrument Science Center (X-ISC) will be provided by an international consortium of ten ESA member states, plus the United States and Japan. Two other European countries (Ireland and United Kingdom) are providing scientific support to X-IFU. 

\begin{figure}[!ht]
    \centering
    \includegraphics[width=17cm]{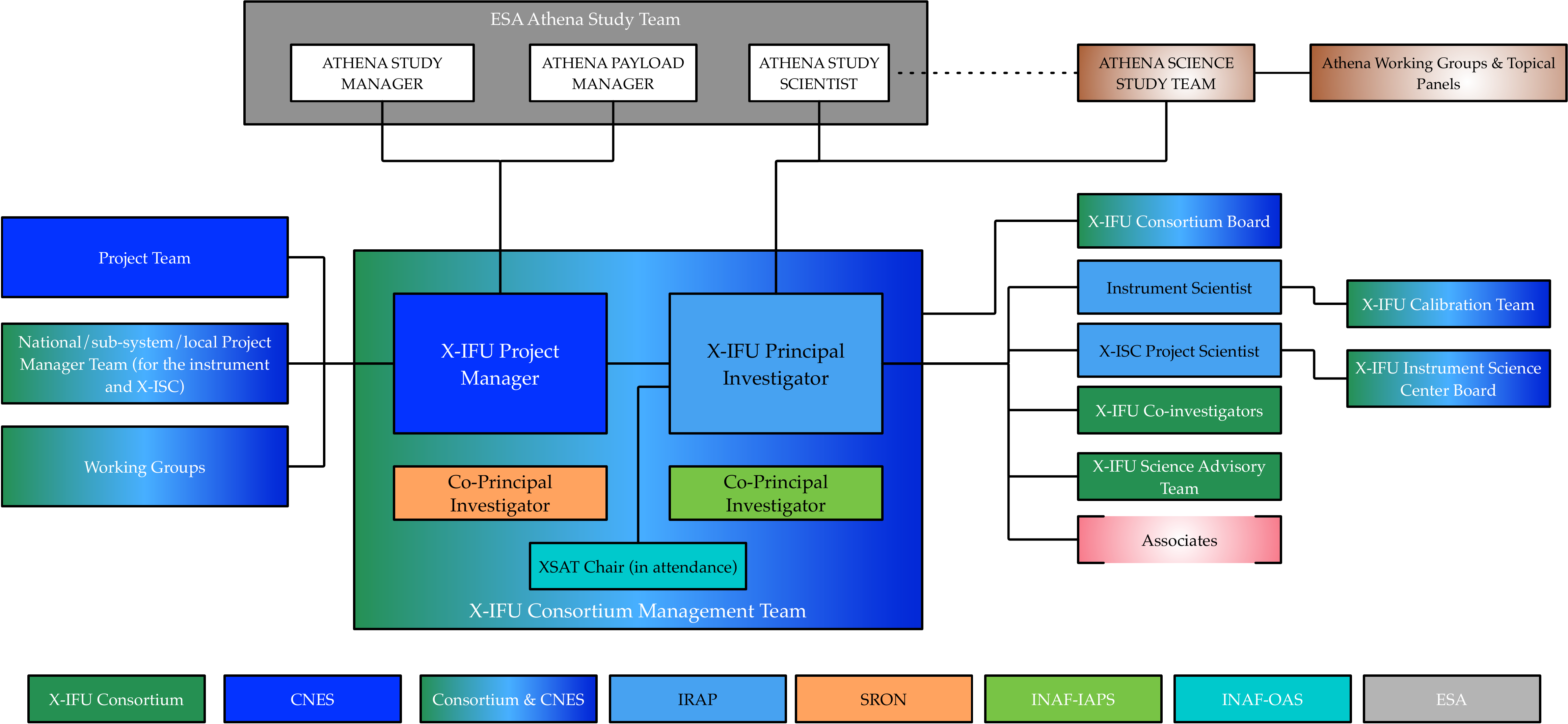}
    \caption{The X-IFU Consortium organisation, highlighting the X-IFU Consortium Management Team, and the Project Manager and Principal Investigator interfaces.}
    \label{fig:organisation}
\end{figure}

The X-IFU Consortium Management Team (XCMT) acts as the governing body of the X-IFU consortium. The XCMT consists of the X-IFU Project Manager (V. Albouys), the X-IFU Principal Investigator (D. Barret), and the two X-IFU Co-Principal Investigators (J.W den Herder and L. Piro), with the X-IFU Science Advisory Team Chair in attendance (M. Cappi). The XCMT is jointly chaired by the CNES PM and the PI, with the PI in charge of organizing the meetings and the day-to-day communication of the XCMT. The XCMT assists and supports the X-IFU PM and X-IFU PI in all X-IFU related matters (technical, programmatic, organizational, scientific). 


The PI and the PM interact with various bodies within the Consortium and with the ESA study team (see Figure \ref{fig:organisation}). This organisation was formally endorsed by ESA in December 2018 through the instrument consortia consolidation process. The X-IFU PI is supported by the X-IFU Science Advisory Team (XSAT), acting as a forum to address all scientific matters related to the development of the X-IFU. The X-IFU PI relies on an X-IFU Calibration Team, chaired by the X-IFU Instrument scientist, to carry out all activities related to the calibration of the instrument, both on-ground and in-flight. The X-IFU Instrument Science Center is the joint responsibility of the X-IFU PM and PI, and its development is delegated to an X-ISC system lead and an X-IFU project scientist. The X-IFU PM prime responsibility are the engineering activities of the X-IFU and the delivery of the instrument and its X-ISC to ESA on behalf of the X-IFU Consortium. The X-IFU PM is the interface with ESA and the industrial primes on aspects related to the interfaces between X-IFU and the Athena satellite. The X-IFU PM is supported by a team of project managers providing a forum for discussing project related issues (e.g. planning). The X-IFU PM is also supported by CNES/Consortium Working Groups. 
 
\begin{figure}
    \centering
    \includegraphics[width=17cm]{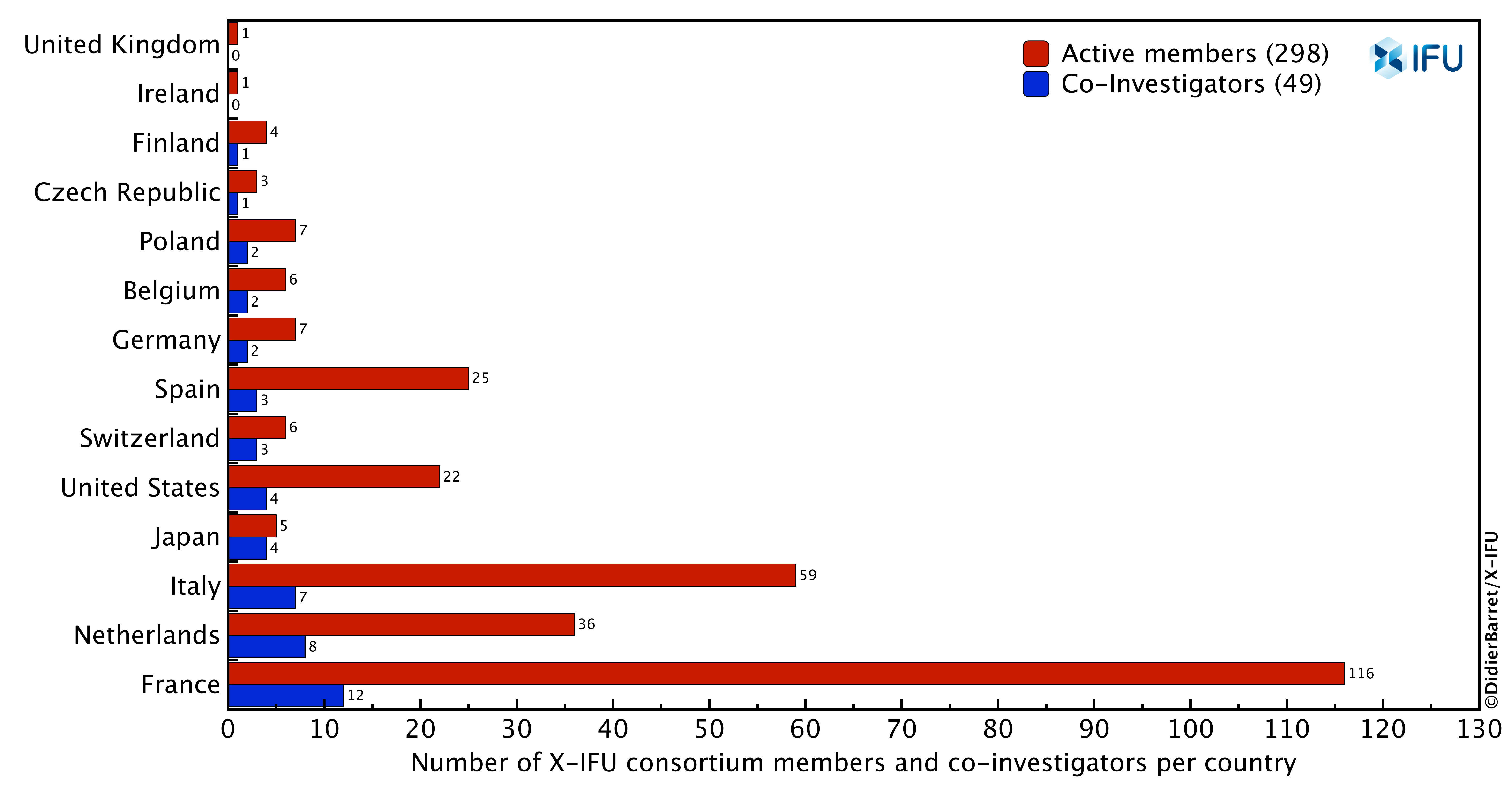}
    \caption{The split of active members of the X-IFU Consortium and X-IFU co-investigators per country. Ireland and the United Kingdom have no identified contributions to the instrument or the ground segment but contribute to scientific activities. None have co-investigators. Note that people identified as contractors, observers or associates, in the Consortium database are not included as active members of the X-IFU Consortium.}
    \label{fig:coi_split}
\end{figure}

Within the current perimeter of the Consortium, the split of X-IFU co-investigators and consortium members is presented in Figure \ref{fig:coi_split}. This gives an overview of the respective efforts put in the X-IFU by each country.
\section{Communication and outreach}
The X-IFU internal communication is managed through various tools, such as the Document Management System CapLinked, Teamwork for action management, and Formsite for online forms and surveys. The external communication is achieved through a dedicated web site\footnote{\url{http://x-ifu.irap.omp.eu}}, which is the gate for accessing X-IFU related resources, e.g. response files. The web site is providing access to outreach material and links to the social media, e.g. a YouTube channel dedicated to X-IFU in which movies explaining X-IFU are available in several languages. Outreach activities are coordinated with the Athena community office\cite{Martinez_2020sea..confE.232M}. The Consortium database, more than 300 entries, is managed through Python, interfacing with the web site (where the various bodies of the Consortium are listed).

\section{X-IFU and its environmental footprint}
As shown above, the X-IFU is a rather complex instrument, involving very advanced technologies, large resources, and a large number of partners spread all across the world. X-IFU is being developed right at the time when it is becoming very clear that without an immediate and drastic reduction of our emissions (by at least 50\% by 2030), across all sectors, the objective of limiting global warming to 1.5 degrees will not be reached. X-IFU is also developed when the need for building new large scientific infrastructures having major environmental impacts is being debated within the community\cite{Knodlseder_2022NatAs...6..503K}, this very same community that should lead by example \cite{Burtscher_2022NatAs...6..764B}. In this context, a life cycle assessment (LCA) of X-IFU is being performed to evaluate the global footprint of developing X-IFU along its complete life cycle. The LCA will identify the contribution of all life cycle phases (from cradle to grave) to several environmental impacts, including global warming (in particular CO$_2$ emissions), waste, water consumption, raw materials consumption, human health, and biodiversity. The idea is simply to identify hot spots and take corrective actions whenever possible, while not jeopardizing the ultimate success of the project. X-IFU has already committed to reduce significantly its global travel footprint\cite{Barret_2020ExA....49..183B}, starting even in the pre-covid era, but the LCA will enable us to go one step further by identifying those actions in our daily activities that can lead to substantial reductions of our impact, such as switching off test equipments over night, and sharing resources and infrastructures wherever possible. At the time of this writing, the assessment has not yet been completed, but the data have been collected from all sub-systems making up the X-IFU. There has been enthusiastic support for the tedious data collection, which is now extrapolated when partial and properly modeled with a dedicated LCA software. There is great hope that the quality of the assessment will enable clear guidelines to be formulated for building the X-IFU, while lowering its environmental impact.

\section{From the SRR onward}
The discovery of an unanticipated cost increase of Athena by the ESA executive now imposes us the challenge of a design-to-cost exercise to bring the cost of Athena within an envelope affordable by the ESA science program ($\sim 1.3$ G€). This exercise should preserve the flagship capabilities of the Athena mission. The X-IFU team, including its international partners, has the expertise and is committed to enter this exercise in a pro-active way, focusing first on the way to simplify the cooling chain. Let alone for feeling accountable for the tax-payer money already spent, the scientific and technical expertise developed over the years should not be wasted, and will be a critical input of the design-to-cost exercise, which will start from the X-IFU design presented at the SRR.

The extent by which X-IFU will change will also depend on the changes applying to the overall telescope and spacecraft, which are under ESA responsibility. A close collaboration between ESA and the instrument teams will be required for Athena to retain the capability to reach most, if not all, the scientific objectives spelled out for the Hot and Energetic Universe, and whose priorities have been repeated constantly since its original formulation in 2013.

\section{Conclusions}
X-IFU entered its system requirement review with a consolidated design and performance budgets understood. As an indication of the extent of the work done for the SRR, the data pack produced by the team consisted of over 100 documents and of more than 5000 pages. This paper aimed at summarizing this documentation, providing a reference point for the upcoming design-to-cost exercise. We, as a team, will use all our knowledge for X-IFU to maintain flagship capabilities in spatially resolved high resolution X-ray spectroscopy, as to enable most of the original X-IFU related scientific objectives of the Athena mission to be retained. 
\label{sec:intro}  

\section{Acknowledgments}
This paper is based on the documentation assembled by the CNES project team and the Consortium partners for the X-IFU System Requirement Review, which has started at the end of June, albeit with reduced objectives (the review will be completed in September 2022). DB wishes to thank the CNES Project and Support teams for continuing the preparation of the SRR data pack, during the disruptive and unfortunate events that happened to Athena, which could have led to the termination of X-IFU. There is no doubt that their efforts will be rewarded in the upcoming design-to-cost exercise. DB would also like to express his gratitude to the CNES Management (Philippe Baptiste, President), Lionel Suchet (Chief Operating Officer), Caroline Laurent (Director of orbital systems and applications), Philippe Lier (Deputy-Director of orbital systems and applications), the French ESA SPC delegation (Olivier La Marle and Juliette Lambin) for their unfailing support to X-IFU. This support was instrumental in preserving an X-IFU on the new Athena mission and will remain precious in the upcoming phase of the reformulation of the mission. Special thanks to all the SPC delegations which supported the objective of keeping a flagship X-ray observatory in the ESA Science Program. DB is also thankful to  Gilles Bergametti (President of the CNES Comité des Programmes Scientifiques, CPS),  Athena Coustenis (President of the CNES Comité d'évaluation de la recherche et de l'exploration spatiale, CERES) and Pierre-Olivier Petrucci (President of the Programme National des Hautes Energies) for their support and for providing the feedback from the French scientific community at large. The French contribution to X-IFU is funded by CNES, CNRS and CEA. This work has been also supported by ASI (Italian Space Agency) through the Contract 2019-27-HH.0, and by the ESA (European Space Agency) Core Technology Program (CTP) Contract No. 4000114932/15/NL/BW and the AREMBES - ESA CTP No.4000116655/16/NL/BW. This publication is part of grant RTI2018-096686-B-C21 funded by MCIN/AEI/ 10.13039/501100011033 and by “ERDF A way of making Europe”. This publication is part of grant RTI2018-096686-B-C21 and PID2020-115325GB-C31 funded by MCIN/AEI/10.13039/501100011033.

We wish to thank an anonymous referee for a very careful reading of our manuscript, and for providing helpful comments. 
\section{Declarations}
There is no conflict of interest.
\bibliography{xifu_biblio_at_ISRR.bib} 
\bibliographystyle{spiebib} 
\end{document}